\documentclass[useAMS,usenatbib]{mn2e}
\usepackage{times}
\usepackage{epsfig}
\usepackage{graphicx}
\usepackage{latexsym}
\usepackage{pifont}
\input epsf

\newcommand{\xmm}{{\it XMM~\/}}
\newcommand{\xmmn}{{\it XMM-Newton~\/}}

\newcommand{\chandra}{{\it Chandra~\/}}

\newcommand{\rosat}{{\it ROSAT~\/}}
\newcommand{\einstein}{{\it Einstein~\/}}

\def\ergcms{{\rm ~erg~cm^{-2}~s^{-1}}}

\def\ergsec{{\rm ~erg~s^{-1}}}
\def\cms{{\rm ~cm^{-2}}}

\def\H0{{\rm ~km~s^{-1}~Mpc^{-1}}}

\def\la{\mathrel{\hbox{\rlap{\hbox{\lower4pt\hbox{$\sim$}}}{\raise2pt\hbox{$<$}}}}}
\def\ga{\mathrel{\hbox{\rlap{\hbox{\lower4pt\hbox{$\sim$}}}{\raise2pt\hbox{$>$}}}}}
\def\d25{D$_{25}$}

\def\hi{H{\small I}$~$}
\def\hii{H{\small II}$~$}

\def\eps@scaling{1.0}%
\newcommand\plottwo[2]{%
  \centering
  \leavevmode
  \columnwidth=.45\columnwidth
  \includegraphics[width={\eps@scaling\columnwidth}]{#1}%
  \hfil
  \includegraphics[width={\eps@scaling\columnwidth}]{#2}%
}%

\title[An \xmmn view of M101 - II] {An \xmmn view of M101 - II. Global X-ray source properties}
\author[L. Jenkins et al.]
	{L.P.\ Jenkins$^1$\thanks{E-mail: lej@star.le.ac.uk}, T.P.\ Roberts$^1$, R.S.\ Warwick$^1$, R.E.\ Kilgard$^{1,2}$, M.J.\ Ward$^1$\\  
$^1$ X-ray \& Observational Astronomy Group, Dept. of Physics \& Astronomy, University of Leicester, University Road, Leicester LE1 7RH, U.K.\\
$^2$ Harvard-Smithsonian Center for Astrophysics, 60 Garden Street, Cambridge, MA 02138, USA. \\}

\date{Accepted ......................; Received .....................; in original form .....................}

\pagerange{\pageref{firstpage}--\pageref{lastpage}}

\pubyear{2004}

\begin{document}

\maketitle

\label{firstpage}

\begin{abstract}

We present the global X-ray properties of the point source population in the grand-design spiral galaxy M101, as seen with {\it XMM-Newton}. 108 X-ray sources are detected within the \d25 ellipse of M101, of which $\sim$ 24 are estimated to be background sources. Multiwavelength cross-correlations show that 20 sources are coincident with \hii regions and/or supernova remnants (SNRs), 7 have identified/candidate background galaxy counterparts, 6 are coincident with foreground stars and one has a radio counterpart. While the spectral and timing properties of the brightest sources were presented in \citet{jenkins04}, here we apply an X-ray colour classification scheme to split the entire source population into different types, i.e. X-ray binaries (XRBs), SNRs, absorbed sources, background sources and supersoft sources (SSSs). Approximately 60 per cent of the population can be classified as XRBs, although there is source contamination from background AGN in this category as they have similar spectral shapes in the X-ray regime. Fifteen sources have X-ray colours consistent with SNRs, three of which correlate with known SNR/\hii radio sources. Another two are promising new candidates for SNRs, one is unidentified, and the remainder are a mixture of foreground stars, bright soft XRBs and AGN candidates. We also detect 14 candidate SSSs, with significant detections in the softest X-ray band (0.3--1\,keV) only. Sixteen sources display short-term variability during the \xmmn observation, twelve of which fall into the XRB category, giving additional evidence of their accreting nature. Using archival \chandra \& \rosat HRI data, we find that $\sim$ 40 per cent of the \xmm sources show long-term variability over a baseline of up to $\sim$ 10 years, and eight sources display potential transient behaviour between observations. Sources with significant flux variations between the \xmm and \chandra observations show a mixture of softening and hardening with increasing luminosity. The spectral and timing properties of the sources coincident with M101 confirm that its X-ray source population is dominated by accreting XRBs.

\end{abstract}


\section{Introduction}
\label{sec:intro}

Since the launch of the \einstein X-ray observatory in the late 1970's, we have known that discrete, point-like X-ray sources form a major constituent of the overall X-ray output of spiral galaxies, along with hot diffuse gas and in some cases an active galactic nucleus (AGN) (e.g. \citealt{fabbiano89}). There are a variety of compact sources that emit with sufficient luminosity in the X-ray regime to be observable in nearby galaxies. These include accreting sources, such as black hole/neutron star X-ray binaries (XRBs) and supersoft sources (SSSs), plus bright thermal supernova remnants (SNRs).

While {\it Einstein}, and later {\it ROSAT}, provided a preliminary view of the X-ray source populations of nearby galaxies (e.g. \citealt{fabbiano87}; \citealt*{read97}; \citealt{roberts00}), these studies were hindered by a lack of both spatial resolution and photon collecting area, leading to source confusion and limited spectroscopic capabilities. Now, with the current generation of X-ray observatories, \xmmn and {\it Chandra}, substantial progress is being made in this field. The sub-arcsecond spatial resolution of \chandra has led to up to a factor 100 increase in the number of sources detected in many galaxies. For example, the recent \chandra survey of 11 nearby spirals of \citet{kilgard04} has resolved on average 75 point sources per galaxy, with luminosities extending down to 5$\times10^{36} \ergsec$. The luminosity functions derived from \chandra studies have also demonstrated that the overall X-ray emission in spiral/star-forming galaxies are typically dominated by a few bright extra-nuclear sources (e.g. \citealt{kilgard02}; \citealt{colbert04}), the brightest of which are the ultraluminous X-ray sources (ULXs) with luminosities $\geq10^{39} \ergsec$ (see \citealt{miller04} for a recent review). The complementary high throughput and large $\sim$ 30 arcminute field-of-view of \xmmn can be used for detailed spectral and variability studies of the brightest point sources in nearby galaxies. This has been demonstrated to great effect in the first paper in this series (\citealt{jenkins04}, hereafter Paper I), as well as the studies of M33 \citep{pietsch04} and the archetypal starburst galaxy NGC~253 \citep{pietsch01}.

The target for this study is M101, a nearby ($d$=7.2\,Mpc, \citealt{stetson98}) grand-design spiral galaxy similar in morphological type to our own Milky Way. Its full face-on aspect and relatively low line-of-sight Galactic hydrogen column ($N_H\sim1.2\times10^{20} \cms$, \citealt{dickey90}) makes it an ideal laboratory for the study of the X-ray emission from compact discrete sources in spiral galaxies.  M101 was first studied at X-ray energies with \einstein (\citealt{mccammon84}; \citealt*{trinchieri90}), revealing X-ray emission associated with the nuclear region and \hii regions in the spiral arms. Subsequent \rosat studies revealed the presence of numerous discrete sources \citep*{wangetal99} as well as a substantial diffuse component \citep{snowden95}. More recently, a \chandra observation has revealed $>$ 100 discrete sources in the central $\sim$8 arcminutes of the galaxy (\citealt{pence01}; \citealt{mukai03}), as well as diffuse emission tracing the spiral arms \citep{kuntz03}. 

This is the second of a series of papers on the \xmmn observation of M101. In Paper I, we presented the spectral and timing properties of the most luminous X-ray sources at the time of the \xmmn observation, many of which were in the ULX regime, and showed properties consistent with high-state XRBs. In this paper, we present the global properties of the complete set of X-ray sources detected within the spatial extent of M101 in the \xmmn observation, complemented with an analysis of two archival sets of \chandra observations of M101. An analysis of the diffuse X-ray emission will be presented in Paper III (Warwick et al., {\it in preparation}). 

This paper is set out as follows. In section~\ref{sec:obs} we outline the observation details and data reduction techniques. In section~\ref{sec:src_detection} we detail the source detection techniques used, and the full source list is presented in section~\ref{sec:catalogue}. In section~\ref{sec:xcolours} we broadly classify the sources according to their X-ray colours, and in section~\ref{sec:timing} we search for short-term variability within the \xmmn observation, and long-term source variability using archival X-ray observations. In section~\ref{sec:class}, we discuss the properties of the various source types, and in section~\ref{sec:conc} we summarise our results.

\section{Observations \& Data Reduction}
\label{sec:obs}

M101 was observed with \xmmn for 42.8\,ks on the 4th June 2002 (ObsID 0104260101). The EPIC MOS-1, MOS-2 \& PN cameras were operated with medium filters in ``Prime Full Window'' mode, which utilizes the full $\sim30$ arcminute field of view of {\it XMM-Newton}, covering the entire \d25 ellipse of M101 ($\sim$28.8 arcminutes diameter, \citealt{devaucouleurs91}). The data were pipeline-processed using the {\small SAS} ({\it Science Analysis Software}) {\small v5.3.2} and all subsequent data analysis was carried out using {\small SAS v5.4.1}. 

In this study we have performed source detection on both the PN and MOS data (see section~\ref{sec:src_detection}). To increase the sensitivity of the MOS data, the MOS1 and MOS2 event lists were merged together using the {\small SAS} task {\small MERGE}. For the purposes of multi-band source detection (see section~\ref{sec:src_detection}), images were created in the following three energy bands: 0.3--1\,keV (soft), 1--2\,keV (medium) and 2--6\,keV (hard). The data were filtered using patterns corresponding to single \& double pixel events for the PN (0 \& 1--4), and patterns 0-12 (single to quadruple events) for the MOS cameras, together with the {\tt \#XMMEA\_EM} (MOS) and {\tt \#XMMEA\_EP} (PN) flags to remove hot pixels or out-of-field events. Full-field lightcurves were accumulated for the three exposures to check for high background intervals of soft proton flares. There were numerous small flares throughout the exposure. For the MOS data, we screened out time intervals corresponding to the four most prominent peaks with count rates greater than $\sim$15 counts per second, leaving a net good time for each camera of 36.7\,ks. In the case of the PN data, inspection of the hard band (2--6\,keV) images showed a greater contamination from flaring. In order to maximise the sensitivity of the data to the source detection algorithms, we further cleaned the data by selecting good time intervals (GTIs) corresponding to less than 0.9 counts per second (10--15\,keV) in the single pattern lightcurve, leaving a net good time for the PN data of 25.7\,ks.  

As in Paper I, this dataset is supplemented with two archival \chandra observations of M101. These consist of one long ($\sim$ 100\,ks) observation performed on the 26th March 2000 (ObsID 934), plus a short ($\sim$ 10\,ks) observation performed on the 29th October 2000 (ObsID 2065). Though the first analysis of the central ACIS-S3 chip in the long observation is presented in \citet{pence01}, we have used the analysis of both data sets presented in \citet{kilgard04} to search for long-term spectral and flux source changes. This analysis utilizes more up-to-date calibration files and includes sources from five ACIS chips, covering a larger fraction (78 per cent) of the \d25 ellipse of M101.

\begin{figure*}
\begin{center}
\includegraphics[width=12cm, height=11.3cm]{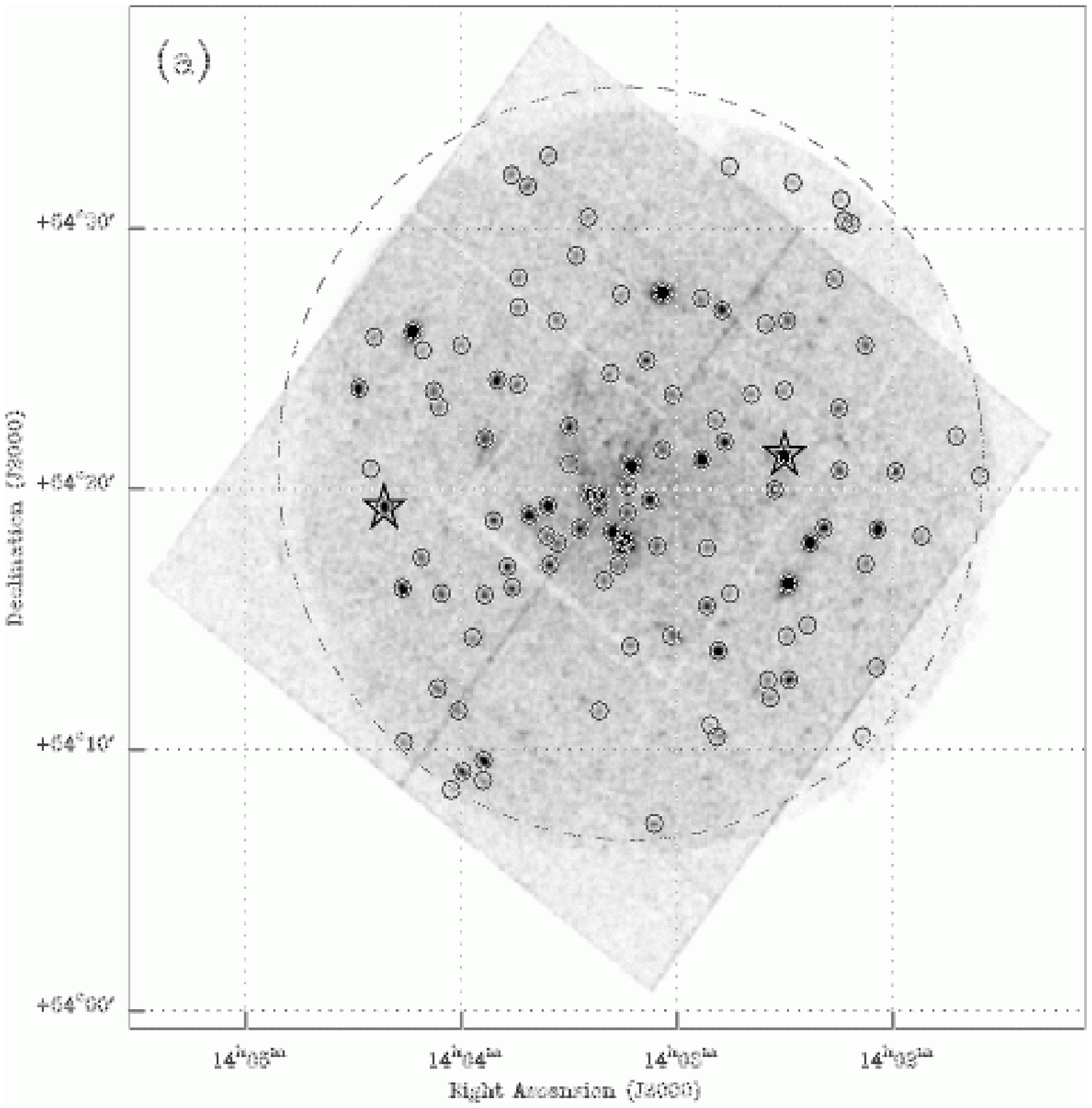}
\includegraphics[width=12cm, height=11.3cm]{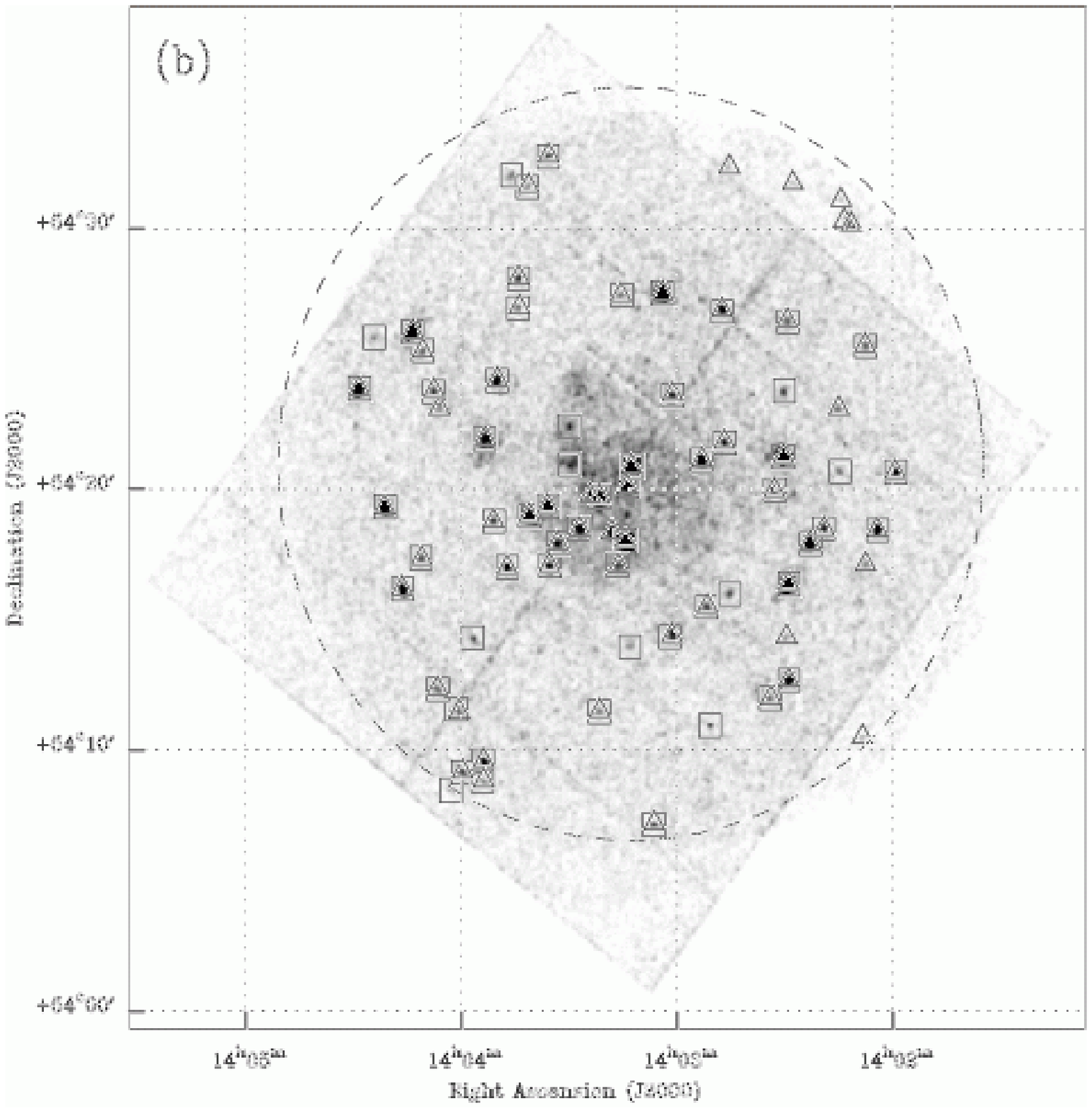}
\end{center}
\caption{\xmm EPIC images of the M101 field (stacked MOS \& PN). (a) full sourcelist overlaid on a broad-band (0.3--6\,keV) image.  The two bright foreground stars in the field (\#25 \& 105) are marked with open star symbols. (b) soft (0.3--1\,keV) image with significant soft sources. PN and MOS detections are denoted with squares and triangles respectively. The \d25 circle is shown with a dashed line.}
\label{fig:xmm}
\end{figure*}

\begin{figure*}
\begin{center}
\includegraphics[width=12cm, height=11.3cm]{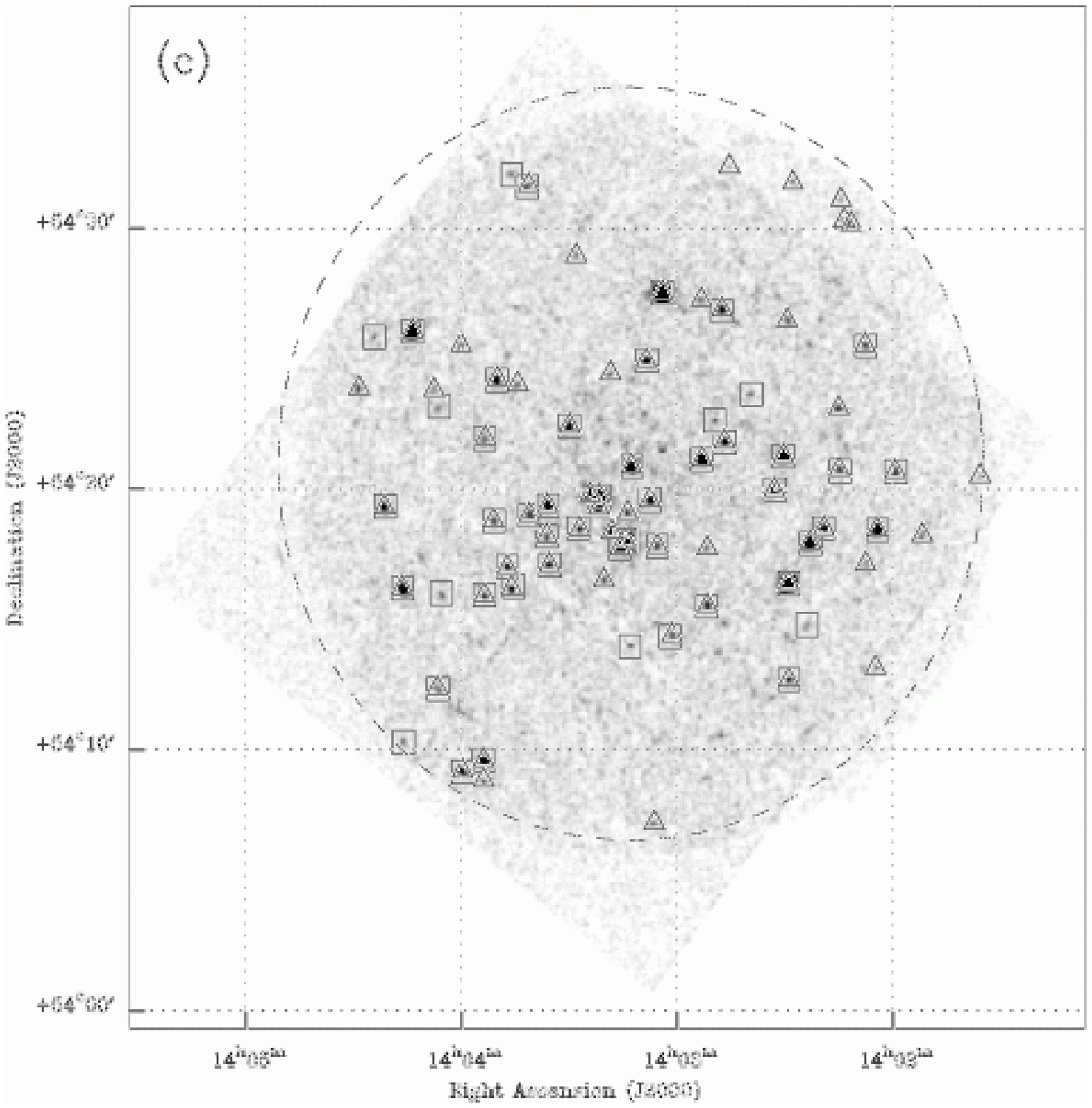}
\includegraphics[width=12cm, height=11.3cm]{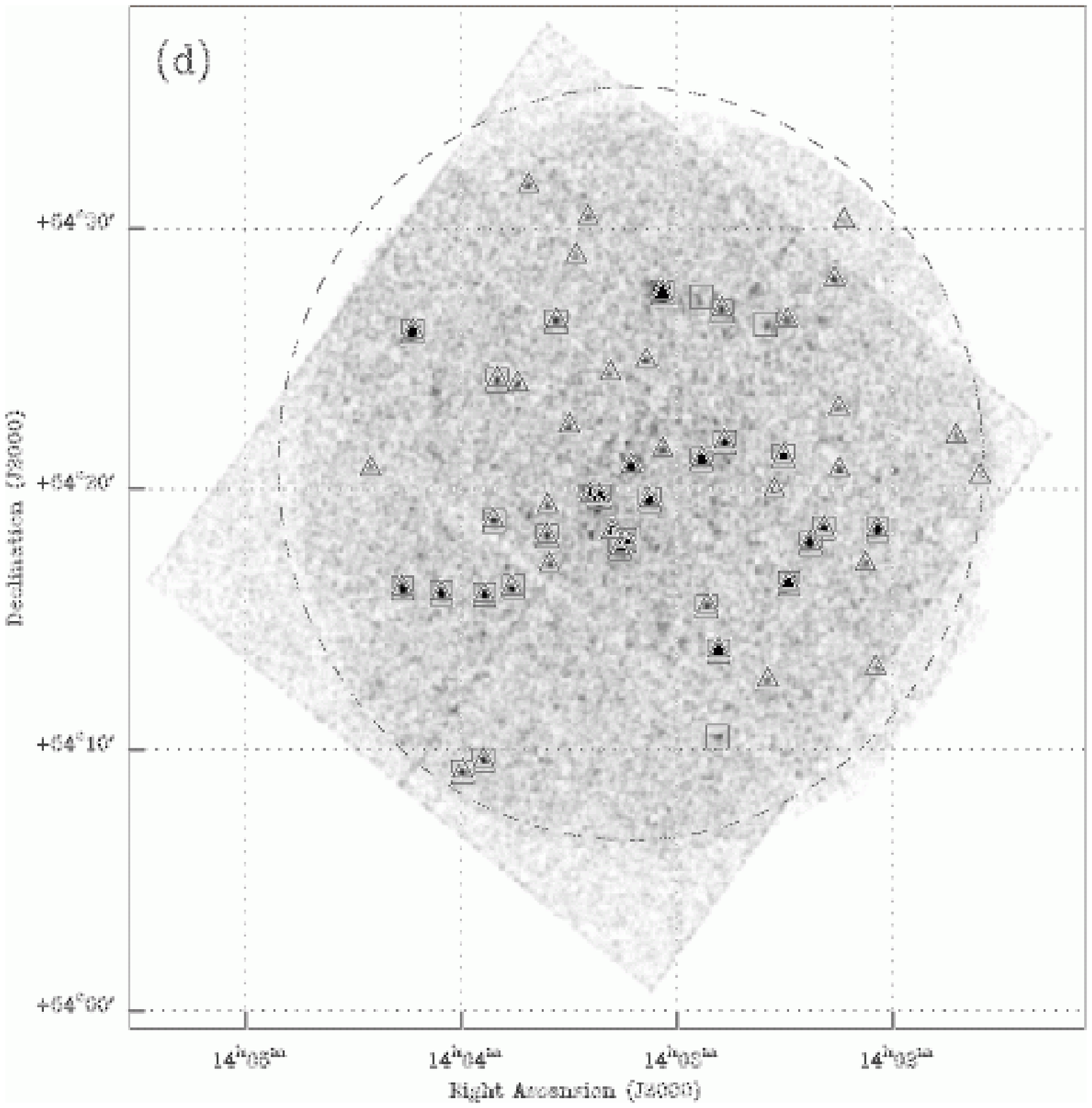}
\end{center}
\contcaption{(c) medium (1--2\,keV) image with significant medium detections. (d) hard (2--6\,keV) image with significant hard detections. PN and MOS detections are denoted with squares and triangles respectively. The \d25 circle is shown with a dashed line.}
\end{figure*}

\begin{table*}
\caption{M101 Source Catalogue.}
 \centering
{\scriptsize
  \begin{tabular}{@{}lccccccccccccc@{}}
\hline

Src & XMMU         & r$_{1\sigma}$ & \multicolumn{3}{c}{PN count rate (count ks$^{-1}$)} & \multicolumn{3}{c}{MOS count rate (count ks$^{-1}$)} & F$_X$   & L$_X$      & HR1            & HR2   & Var \\
    &                  & ($^{\prime\prime}$)   & S      & M            & H            & S            & M            & H            &    &                   &          &     & \\
(1) & (2)              & (3)    & \multicolumn{3}{c}{(4)}              & \multicolumn{3}{c}{(5)}                    & (6)            & (7)            & (8)            & (9)  & (10) \\

\hline

1   & J140134.7+542031 & 2.05 & -            & -            & -            & 0.3$\pm$0.2  & {\bf 1.3$\pm$0.2}  & {\bf 1.1$\pm$0.3}  & 2.28$\pm$0.41  & 1.42$\pm$0.25  & 0.59$\pm$0.18  & -0.09$\pm$0.15 \\
2   & J140141.3+542202 & 2.43 & -            & -            & -            & 0.3$\pm$0.2  & 0.5$\pm$0.2  & {\bf 1.0$\pm$0.2}  & 1.72$\pm$0.34  & 1.07$\pm$0.21  & 0.25$\pm$0.33  & 0.34$\pm$0.20 \\
3   & J140151.1+541814 & 2.20 & 0.6$\pm$0.5  & 2.0$\pm$0.6  & 1.4$\pm$0.7  & 0.1$\pm$0.1  & {\bf 0.9$\pm$0.2}  & 0.2$\pm$0.2  & 0.87$\pm$0.21  & 0.54$\pm$0.13  & 0.75$\pm$0.17  & -0.41$\pm$0.18 & L\\
4   & J140158.4+542042 & 1.72 & {\bf 7.0$\pm$1.0}  & {\bf 3.4$\pm$0.7}  & 0.9$\pm$0.5  & {\bf 0.9$\pm$0.2}  & {\bf 1.0$\pm$0.2}  & 0.6$\pm$0.2  & 1.69$\pm$0.21  & 1.05$\pm$0.13  & -0.18$\pm$0.09 & -0.38$\pm$0.13 \\
5   & J140203.5+541828 & 1.56 & {\bf 9.7$\pm$1.1}  & {\bf 10.8$\pm$1.1} & {\bf 5.7$\pm$0.9}  & {\bf 3.1$\pm$0.3}  & {\bf 3.3$\pm$0.4}  & {\bf 2.8$\pm$0.3}  & 6.34$\pm$0.35  & 3.93$\pm$0.22  & 0.05$\pm$0.05  & -0.18$\pm$0.06 & S,L\\
6   & J140203.9+541312 & 1.89 & 0.5$\pm$0.5  & 1.9$\pm$0.6  & 1.7$\pm$0.7  & 0.3$\pm$0.2  & {\bf 0.7$\pm$0.2}  & {\bf 1.0$\pm$0.2}  & 1.58$\pm$0.26  & 0.98$\pm$0.16  & 0.47$\pm$0.22  & 0.08$\pm$0.14 \\
7   & J140206.8+542532 & 1.73 & {\bf 3.0$\pm$0.7}  & {\bf 2.1$\pm$0.6}  & 0.8$\pm$0.5  & {\bf 0.7$\pm$0.1}  & {\bf 0.8$\pm$0.2}  & 0.4$\pm$0.2  & 1.20$\pm$0.19  & 0.75$\pm$0.12  & -0.01$\pm$0.11 & -0.37$\pm$0.16 \\
8   & J140206.9+541710 & 1.92 & 1.8$\pm$0.5  & 1.2$\pm$0.5  & 1.4$\pm$0.5  & {\bf 0.6$\pm$0.1}  & {\bf 0.7$\pm$0.2}  & {\bf 0.8$\pm$0.2}  & 1.50$\pm$0.19  & 0.93$\pm$0.12  & -0.04$\pm$0.13 & 0.11$\pm$0.13 \\
9   & J140207.9+541033 & 2.14 & -            & -            & -            & {\bf 1.1$\pm$0.2}  & 0.8$\pm$0.2  & 0.7$\pm$0.3  & 1.75$\pm$0.43  & 1.08$\pm$0.26  & -0.21$\pm$0.19 & -0.06$\pm$0.27 \\
10  & J140210.6+543011 & 1.72 & -            & -            & -            & {\bf 3.5$\pm$0.4}  & {\bf 1.7$\pm$0.3}  & 0.2$\pm$0.2  & 2.58$\pm$0.36  & 1.60$\pm$0.22  & -0.34$\pm$0.09 & -0.78$\pm$0.19 \\
11  & J140212.6+543019 & 2.04 & -            & -            & -            & {\bf 1.0$\pm$0.2}  & {\bf 1.2$\pm$0.2}  & {\bf 1.1$\pm$0.2}  & 2.43$\pm$0.35  & 1.51$\pm$0.22  & 0.10$\pm$0.15  & -0.04$\pm$0.13 \\
12  & J140213.5+543107 & 2.10 & -            & -            & -            & {\bf 0.7$\pm$0.2}  & {\bf 0.7$\pm$0.2}  & 0.5$\pm$0.2  & 1.30$\pm$0.35  & 0.81$\pm$0.22  & 0.03$\pm$0.18  & -0.20$\pm$0.27 \\
13  & J140214.1+542045 & 1.81 & {\bf 2.1$\pm$0.5}  & {\bf 2.3$\pm$0.5}  & 1.9$\pm$0.5  & 0.3$\pm$0.1  & {\bf 0.6$\pm$0.1}  & {\bf 0.9$\pm$0.2}  & 1.61$\pm$0.19  & 1.00$\pm$0.12  & 0.18$\pm$0.12  & 0.08$\pm$0.11 & L \\
14  & J140214.3+542308 & 1.86 & -            & -            & -            & {\bf 0.5$\pm$0.1}  & {\bf 1.1$\pm$0.2}  & {\bf 0.6$\pm$0.1}  & 1.49$\pm$0.22  & 0.92$\pm$0.14  & 0.32$\pm$0.12  & -0.29$\pm$0.14 \\
15  & J140215.4+542805 & 1.97 & -            & -            & -            & 0.1$\pm$0.1  & 0.5$\pm$0.2  & {\bf 1.4$\pm$0.2}  & 2.13$\pm$0.32  & 1.32$\pm$0.20  & 0.72$\pm$0.32  & 0.50$\pm$0.14 \\
16  & J140218.6+541832 & 1.66 & {\bf 4.5$\pm$0.7}  & {\bf 2.6$\pm$0.5}  & {\bf 2.3$\pm$0.6}  & {\bf 0.8$\pm$0.1}  & {\bf 1.4$\pm$0.2}  & {\bf 1.3$\pm$0.2}  & 2.51$\pm$0.22  & 1.56$\pm$0.13  & 0.04$\pm$0.08  & -0.04$\pm$0.08 \\
17  & J140222.4+541758 & 1.54 & {\bf 14.6$\pm$1.1} & {\bf 10.2$\pm$0.9} & {\bf 5.7$\pm$0.8}  & {\bf 3.2$\pm$0.3}  & {\bf 3.8$\pm$0.3}  & {\bf 1.9$\pm$0.2}  & 5.91$\pm$0.28  & 3.66$\pm$0.18  & -0.07$\pm$0.04 & -0.31$\pm$0.05 & L\\
18  & J140223.2+541450 & 3.16 & 0.3$\pm$0.4  & {\bf 1.6$\pm$0.4}  & 0.7$\pm$0.4  & -            & -            & -            & 0.68$\pm$0.25  & 0.42$\pm$0.16  & 0.68$\pm$0.33  & -0.40$\pm$0.28 \\
19  & J140227.1+543146 & 1.96 & -            & -            & -            & {\bf 0.9$\pm$0.2}  & {\bf 1.1$\pm$0.2}  & 0.7$\pm$0.2  & 1.80$\pm$0.31  & 1.12$\pm$0.19  & 0.09$\pm$0.14  & -0.22$\pm$0.17 & S\\
20  & J140228.4+541625 & 1.52 & {\bf 22.9$\pm$1.3} & {\bf 34.0$\pm$1.6} & {\bf 13.9$\pm$1.1} & {\bf 5.1$\pm$0.4}  & {\bf 10.4$\pm$0.5} & {\bf 7.1$\pm$0.5}  & 16.03$\pm$0.48 & 9.94$\pm$0.30  & 0.26$\pm$0.03  & -0.31$\pm$0.03 & S,L\\
21  & J140228.5+541244 & 1.61 & {\bf 11.5$\pm$1.1} & {\bf 3.0$\pm$0.6}  & 1.7$\pm$0.6  & {\bf 2.7$\pm$0.3}  & {\bf 1.2$\pm$0.2}  & 0.0$\pm$0.1  & 1.95$\pm$0.18  & 1.21$\pm$0.11  & -0.50$\pm$0.06 & -0.76$\pm$0.11 & L\\
22  & J140228.7+542629 & 1.78 & {\bf 3.5$\pm$0.7}  & 1.2$\pm$0.4  & 2.0$\pm$0.6  & {\bf 0.6$\pm$0.1}  & {\bf 0.8$\pm$0.1}  & {\bf 0.8$\pm$0.2}  & 1.67$\pm$0.20  & 1.04$\pm$0.12  & -0.16$\pm$0.10 & 0.08$\pm$0.12 & L\\
23  & J140229.1+541424 & 2.34 & -            & -            & -            & {\bf 0.6$\pm$0.1}  & 0.5$\pm$0.1  & 0.1$\pm$0.1  & 0.53$\pm$0.15  & 0.33$\pm$0.09  & -0.13$\pm$0.18 & -0.79$\pm$0.34 \\
24  & J140229.5+542349 & 2.07 & {\bf 9.0$\pm$1.8}  & 0.0$\pm$0.7  & 0.2$\pm$0.5  & -            & -            & -            & 1.08$\pm$0.37  & 0.67$\pm$0.23  & -1.00$\pm$0.14 & 1.00$\pm$7.81 & L\\
25  & J140229.8+542118 & 1.51 & {\bf 62.5$\pm$2.1} & {\bf 17.4$\pm$1.1} & {\bf 2.6$\pm$0.5}  & {\bf 12.7$\pm$0.5} & {\bf 6.9$\pm$0.4}  & {\bf 1.3$\pm$0.2}  & 10.70$\pm$0.27 & 6.64$\pm$0.17  & -0.47$\pm$0.02 & -0.70$\pm$0.03 & S,L\\
26  & J140232.4+542001 & 1.68 & {\bf 1.9$\pm$0.5}  & {\bf 2.2$\pm$0.4}  & 1.2$\pm$0.4  & {\bf 0.6$\pm$0.1}  & {\bf 1.0$\pm$0.1}  & {\bf 0.5$\pm$0.1}  & 1.35$\pm$0.14  & 0.84$\pm$0.09  & 0.17$\pm$0.09  & -0.28$\pm$0.10 \\
27  & J140233.6+541203 & 1.97 & {\bf 2.8$\pm$0.6}  & 0.8$\pm$0.4  & 0.8$\pm$0.5  & {\bf 0.8$\pm$0.2}  & 0.6$\pm$0.2  & 0.6$\pm$0.2  & 1.09$\pm$0.20  & 0.68$\pm$0.13  & -0.36$\pm$0.14 & 0.00$\pm$0.21 \\
28  & J140234.3+541244 & 2.33 & -            & -            & -            & 0.2$\pm$0.1  & 0.5$\pm$0.2  & {\bf 1.0$\pm$0.2}  & 1.69$\pm$0.29  & 1.05$\pm$0.18  & 0.52$\pm$0.28  & 0.29$\pm$0.17 \\
29  & J140234.8+542622 & 3.42 & 0.3$\pm$0.4  & 0.0$\pm$0.1  & {\bf 3.4$\pm$0.9}  & -            & -            & -            & 1.92$\pm$0.48  & 1.19$\pm$0.30  & -1.00$\pm$0.96 & 1.00$\pm$0.08 & L\\
30  & J140238.8+542340 & 2.15 & 1.1$\pm$0.4  & {\bf 1.4$\pm$0.3}  & 1.1$\pm$0.4  & -            & -            & -            & 0.94$\pm$0.23  & 0.58$\pm$0.14  & 0.14$\pm$0.21  & -0.13$\pm$0.21 & L\\
31  & J140244.9+541602 & 2.09 & {\bf 3.0$\pm$0.5}  & 0.1$\pm$0.2  & 0.0$\pm$0.3  & -            & -            & -            & 0.33$\pm$0.16  & 0.21$\pm$0.10  & -0.96$\pm$0.11 & -1.00$\pm$9.59 \\
32  & J140244.9+543224 & 2.26 & -            & -            & -            & {\bf 0.9$\pm$0.2}  & {\bf 0.7$\pm$0.2}  & 0.1$\pm$0.1  & 0.85$\pm$0.21  & 0.53$\pm$0.13  & -0.16$\pm$0.16 & -0.72$\pm$0.30 & L\\
33  & J140246.4+542151 & 1.60 & {\bf 3.1$\pm$0.5}  & {\bf 2.6$\pm$0.4}  & {\bf 2.7$\pm$0.5}  & {\bf 0.7$\pm$0.1}  & {\bf 1.1$\pm$0.1}  & {\bf 1.0$\pm$0.1}  & 2.21$\pm$0.17  & 1.37$\pm$0.10  & 0.08$\pm$0.08  & -0.02$\pm$0.07 \\
34  & J140247.0+542653 & 1.60 & {\bf 19.4$\pm$3.2} & {\bf 11.2$\pm$2.4} & {\bf 8.2$\pm$2.4}  & {\bf 1.5$\pm$0.2}  & {\bf 1.0$\pm$0.2}  & {\bf 0.6$\pm$0.1}  & 2.07$\pm$0.21  & 1.28$\pm$0.13  & -0.23$\pm$0.08 & -0.20$\pm$0.10 & L\\
35  & J140248.2+541351 & 1.62 & 0.6$\pm$0.4  & 1.1$\pm$0.3  & {\bf 8.0$\pm$0.9}  & 0.1$\pm$0.1  & 0.2$\pm$0.1  & {\bf 3.5$\pm$0.3}  & 4.81$\pm$0.34  & 2.99$\pm$0.21  & 0.32$\pm$0.24  & 0.82$\pm$0.05 & L\\
36  & J140248.5+541033 & 2.79 & 1.3$\pm$0.5  & 1.9$\pm$0.6  & {\bf 2.6$\pm$0.7}  & -            & -            & -            & 1.90$\pm$0.38  & 1.18$\pm$0.23  & 0.20$\pm$0.24  & 0.17$\pm$0.19 \\
37  & J140248.9+542242 & 1.98 & 0.1$\pm$0.2  & {\bf 1.1$\pm$0.3}  & 0.6$\pm$0.3  & 0.1$\pm$0.1  & 0.6$\pm$0.2  & 0.5$\pm$0.2  & 0.62$\pm$0.15  & 0.38$\pm$0.09  & 0.79$\pm$0.22  & -0.18$\pm$0.19 \\
38  & J140250.4+541100 & 1.83 & {\bf 3.6$\pm$0.6}  & 0.0$\pm$0.2  & 0.0$\pm$0.1  & -            & -            & -            & 0.39$\pm$0.09  & 0.25$\pm$0.06  & -1.00$\pm$0.09 & 0.00$\pm$0.00 \\
39  & J140251.3+541534 & 1.66 & {\bf 2.3$\pm$0.5}  & {\bf 2.0$\pm$0.4}  & {\bf 1.5$\pm$0.4}  & {\bf 0.7$\pm$0.1}  & {\bf 1.0$\pm$0.2}  & {\bf 0.8$\pm$0.1}  & 1.63$\pm$0.16  & 1.01$\pm$0.10  & 0.08$\pm$0.09  & -0.12$\pm$0.10 \\
40  & J140251.4+541747 & 2.01 & 1.3$\pm$0.4  & 0.5$\pm$0.3  & 0.7$\pm$0.3  & 0.2$\pm$0.1  & {\bf 0.4$\pm$0.1}  & 0.1$\pm$0.1  & 0.49$\pm$0.11  & 0.30$\pm$0.07  & -0.07$\pm$0.17 & -0.26$\pm$0.21 \\
41  & J140252.8+542111 & 1.54 & {\bf 4.7$\pm$0.6}  & {\bf 5.9$\pm$0.6}  & {\bf 4.3$\pm$0.6}  & {\bf 1.3$\pm$0.2}  & {\bf 2.3$\pm$0.2}  & {\bf 1.7$\pm$0.2}  & 3.88$\pm$0.21  & 2.41$\pm$0.13  & 0.19$\pm$0.06  & -0.15$\pm$0.05 & S,L\\
42  & J140252.9+542721 & 1.94 & 2.1$\pm$1.5  & 3.0$\pm$1.5  & {\bf 8.0$\pm$2.3}  & 0.3$\pm$0.1  & {\bf 0.4$\pm$0.1}  & 0.1$\pm$0.1  & 0.49$\pm$0.15  & 0.30$\pm$0.09  & 0.20$\pm$0.19  & 0.09$\pm$0.18 \\
43  & J140301.0+542340 & 2.07 & {\bf 3.0$\pm$0.6}  & 0.0$\pm$0.1  & 0.0$\pm$0.1  & {\bf 0.6$\pm$0.1}  & 0.0$\pm$0.0  & 0.0$\pm$0.1  & 0.31$\pm$0.07  & 0.19$\pm$0.04  & -1.00$\pm$0.03 & -1.00$\pm$4.95 & L\\
44  & J140301.4+541426 & 2.32 & {\bf 19.8$\pm$2.9} & {\bf 7.6$\pm$1.8}  & 1.1$\pm$1.3  & {\bf 1.2$\pm$0.2}  & {\bf 0.6$\pm$0.1}  & 0.0$\pm$0.0  & 0.81$\pm$0.10  & 0.51$\pm$0.06  & -0.40$\pm$0.08 & -0.95$\pm$0.12 \\
45  & J140303.8+542133 & 2.02 & -            & -            & -            & 0.1$\pm$0.1  & 0.3$\pm$0.1  & {\bf 0.6$\pm$0.1}  & 1.04$\pm$0.18  & 0.65$\pm$0.11  & 0.50$\pm$0.38  & 0.36$\pm$0.16 & L\\
46  & J140303.9+542734 & 1.51 & {\bf 46.0$\pm$1.9} & {\bf 66.4$\pm$2.2} & {\bf 50.9$\pm$2.0} & {\bf 9.7$\pm$0.5}  & {\bf 22.0$\pm$0.7} & {\bf 19.5$\pm$0.7} & 41.98$\pm$0.79 & 26.04$\pm$0.49 & 0.28$\pm$0.02  & -0.09$\pm$0.02 & S,L\\
47  & J140305.2+541751 & 1.78 & 1.5$\pm$0.5  & {\bf 1.8$\pm$0.4}  & 0.5$\pm$0.3  & 0.2$\pm$0.1  & {\bf 0.5$\pm$0.1}  & 0.3$\pm$0.1  & 0.72$\pm$0.11  & 0.45$\pm$0.07  & 0.24$\pm$0.14  & -0.37$\pm$0.14 \\
48  & J140306.2+540714 & 1.86 & {\bf 4.7$\pm$0.9}  & 1.8$\pm$0.6  & 1.5$\pm$0.7  & {\bf 1.5$\pm$0.3}  & {\bf 1.3$\pm$0.3}  & 0.4$\pm$0.3  & 1.75$\pm$0.28  & 1.08$\pm$0.18  & -0.23$\pm$0.11 & -0.34$\pm$0.18 & L\\
49  & J140307.4+541937 & 1.63 & 0.6$\pm$0.3  & {\bf 2.1$\pm$0.4}  & {\bf 3.1$\pm$0.5}  & 0.2$\pm$0.1  & {\bf 0.8$\pm$0.1}  & {\bf 1.4$\pm$0.2}  & 2.29$\pm$0.18  & 1.42$\pm$0.11  & 0.60$\pm$0.13  & 0.23$\pm$0.07 & L\\
50  & J140308.3+542458 & 1.68 & 1.2$\pm$0.4  & {\bf 2.6$\pm$0.4}  & 1.4$\pm$0.4  & 0.2$\pm$0.1  & {\bf 0.8$\pm$0.1}  & {\bf 0.5$\pm$0.1}  & 1.22$\pm$0.13  & 0.76$\pm$0.08  & 0.45$\pm$0.11  & -0.25$\pm$0.10 & L\\
51  & J140312.4+542056 & 1.55 & {\bf 11.7$\pm$1.4} & {\bf 4.9$\pm$0.9}  & 1.6$\pm$0.6  & {\bf 3.2$\pm$0.3}  & {\bf 2.4$\pm$0.2}  & {\bf 1.3$\pm$0.2}  & 3.87$\pm$0.22  & 2.40$\pm$0.14  & -0.22$\pm$0.05 & -0.33$\pm$0.06 & S,L\\
52  & J140312.9+541401 & 2.03 & {\bf 1.8$\pm$0.4}  & {\bf 1.6$\pm$0.4}  & 1.3$\pm$0.4  & -            & -            & -            & 1.18$\pm$0.25  & 0.73$\pm$0.15  & -0.06$\pm$0.17 & -0.09$\pm$0.19 \\
53  & J140313.6+542010 & 1.79 & {\bf 4.9$\pm$0.7}  & 0.0$\pm$0.0  & 0.0$\pm$0.1  & {\bf 1.0$\pm$0.1}  & 0.0$\pm$0.1  & 0.0$\pm$0.0  & 0.47$\pm$0.06  & 0.29$\pm$0.04  & -1.00$\pm$0.02 & 1.00$\pm$23.56 & L\\
54  & J140313.6+541909 & 2.11 & -            & -            & -            & 0.3$\pm$0.1  & {\bf 0.5$\pm$0.1}  & 0.1$\pm$0.1  & 0.57$\pm$0.15  & 0.35$\pm$0.09  & 0.20$\pm$0.20  & -0.55$\pm$0.24 \\
55  & J140314.3+541806 & 1.50 & {\bf 65.8$\pm$2.0} & {\bf 39.2$\pm$0.3} & {\bf 20.1$\pm$1.2} & {\bf 17.7$\pm$0.6} & {\bf 17.2$\pm$0.6} & {\bf 9.7$\pm$0.2}  & 27.44$\pm$0.37 & 17.02$\pm$0.23 & -0.19$\pm$0.01 & -0.29$\pm$0.01 & S,L\\
56  & J140315.4+542730 & 2.13 & {\bf 2.4$\pm$0.5}  & 0.0$\pm$0.1  & 0.2$\pm$0.3  & {\bf 0.4$\pm$0.1}  & 0.2$\pm$0.1  & 0.0$\pm$0.1  & 0.32$\pm$0.10  & 0.20$\pm$0.06  & -0.89$\pm$0.10 & -0.35$\pm$0.62 & L\\
57  & J140315.8+541747 & 1.59 & 0.0$\pm$0.2  & {\bf 2.3$\pm$0.2}  & {\bf 3.0$\pm$0.5}  & 0.0$\pm$0.1  & {\bf 0.7$\pm$0.2}  & {\bf 1.0$\pm$0.1}  & 1.70$\pm$0.13  & 1.05$\pm$0.08  & 1.00$\pm$0.12  & 0.15$\pm$0.08 & L\\
58  & J140316.3+541708 & 2.25 & {\bf 2.4$\pm$0.5}  & 0.2$\pm$0.2  & 0.2$\pm$0.2  & {\bf 0.6$\pm$0.1}  & 0.0$\pm$0.0  & 0.1$\pm$0.1  & 0.39$\pm$0.10  & 0.24$\pm$0.06  & -0.93$\pm$0.11 & 0.51$\pm$0.64 & L\\
59  & J140318.1+541825 & 1.69 & -            & -            & -            & {\bf 1.0$\pm$0.2}  & {\bf 1.1$\pm$0.2}  & {\bf 0.9$\pm$0.1}  & 2.22$\pm$0.23  & 1.38$\pm$0.14  & 0.01$\pm$0.11  & -0.07$\pm$0.11 \\
60  & J140318.5+542429 & 1.79 & -            & -            & -            & 0.3$\pm$0.1  & {\bf 0.6$\pm$0.1}  & {\bf 0.7$\pm$0.1}  & 1.29$\pm$0.17  & 0.80$\pm$0.11  & 0.39$\pm$0.18  & 0.07$\pm$0.12 & L\\
61  & J140320.3+541632 & 2.04 & 0.7$\pm$0.4  & 1.0$\pm$0.3  & 0.4$\pm$0.3  & 0.1$\pm$0.1  & {\bf 0.4$\pm$0.1}  & 0.3$\pm$0.1  & 0.52$\pm$0.12  & 0.32$\pm$0.08  & 0.48$\pm$0.20  & -0.32$\pm$0.19 \\
62  & J140321.4+541133 & 1.95 & {\bf 1.9$\pm$0.5}  & 0.7$\pm$0.4  & 0.0$\pm$0.2  & {\bf 0.9$\pm$0.2}  & 0.4$\pm$0.2  & 0.2$\pm$0.1  & 0.48$\pm$0.13  & 0.30$\pm$0.08  & -0.42$\pm$0.15 & -0.54$\pm$0.34 \\
63  & J140321.6+541946 & 1.52 & {\bf 6.0$\pm$0.8}  & {\bf 12.1$\pm$1.0} & {\bf 6.0$\pm$0.7}  & {\bf 1.6$\pm$0.2}  & {\bf 3.9$\pm$0.1}  & {\bf 2.7$\pm$0.3}  & 6.00$\pm$0.28  & 3.72$\pm$0.17  & 0.39$\pm$0.04  & -0.24$\pm$0.04 & L\\
64  & J140321.7+541920 & 1.59 & -            & -            & -            & 0.4$\pm$0.1  & {\bf 1.2$\pm$0.1}  & 0.4$\pm$0.1  & 1.23$\pm$0.19  & 0.76$\pm$0.12  & 0.52$\pm$0.13  & -0.50$\pm$0.12 \\
65  & J140324.2+541949 & 1.56 & -            & -            & -            & {\bf 1.0$\pm$0.1}  & {\bf 3.4$\pm$0.3}  & {\bf 3.3$\pm$0.3}  & 6.52$\pm$0.41  & 4.05$\pm$0.26  & 0.54$\pm$0.06  & -0.01$\pm$0.06 & S,L\\
66  & J140324.8+543027 & 2.20 & -            & -            & -            & 0.4$\pm$0.1  & 0.4$\pm$0.1  & {\bf 0.8$\pm$0.2}  & 1.41$\pm$0.24  & 0.87$\pm$0.15  & -0.01$\pm$0.26 & 0.34$\pm$0.19 \\
67  & J140327.2+541831 & 1.78 & {\bf 4.3$\pm$0.8}  & {\bf 1.4$\pm$0.4}  & 0.0$\pm$0.2  & {\bf 1.0$\pm$0.2}  & {\bf 0.4$\pm$0.1}  & 0.1$\pm$0.1  & 0.72$\pm$0.10  & 0.45$\pm$0.06  & -0.44$\pm$0.08 & -0.87$\pm$0.18 \\
68  & J140328.1+542859 & 2.49 & -            & -            & -            & 0.3$\pm$0.1  & {\bf 0.5$\pm$0.1}  & {\bf 0.8$\pm$0.2}  & 1.43$\pm$0.24  & 0.89$\pm$0.15  & 0.31$\pm$0.22  & 0.20$\pm$0.16 & L\\
69  & J140329.9+542100 & 2.25 & {\bf 3.1$\pm$0.6}  & 0.2$\pm$0.2  & 0.0$\pm$0.2  & -            & -            & -            & 0.36$\pm$0.12  & 0.22$\pm$0.08  & -0.89$\pm$0.12 & -1.00$\pm$1.92 \\
70  & J140330.0+542228 & 1.68 & {\bf 2.4$\pm$0.5}  & {\bf 2.7$\pm$0.5}  & 0.8$\pm$0.4  & 0.5$\pm$0.1  & {\bf 1.1$\pm$0.2}  & {\bf 0.7$\pm$0.1}  & 1.45$\pm$0.16  & 0.90$\pm$0.10  & 0.24$\pm$0.10  & -0.32$\pm$0.10 & S,L\\
71  & J140333.2+541759 & 1.81 & {\bf 3.5$\pm$0.6}  & 0.0$\pm$0.2  & 0.0$\pm$0.1  & {\bf 0.8$\pm$0.1}  & 0.0$\pm$0.0  & 0.0$\pm$0.0  & 0.37$\pm$0.06  & 0.23$\pm$0.04  & -1.00$\pm$0.07 & 0.00$\pm$0.00 & L\\
72  & J140333.6+542629 & 1.92 & 0.9$\pm$0.7  & 0.9$\pm$0.6  & {\bf 4.2$\pm$1.1}  & 0.0$\pm$0.0  & 0.2$\pm$0.1  & {\bf 0.7$\pm$0.1}  & 1.21$\pm$0.19  & 0.75$\pm$0.12  & 0.60$\pm$0.33  & 0.61$\pm$0.14 \\
73  & J140335.4+541708 & 1.65 & {\bf 3.4$\pm$0.6}  & {\bf 1.9$\pm$0.4}  & 1.4$\pm$0.4  & {\bf 0.7$\pm$0.1}  & {\bf 1.0$\pm$0.2}  & {\bf 0.5$\pm$0.1}  & 1.45$\pm$0.15  & 0.90$\pm$0.09  & -0.03$\pm$0.09 & -0.25$\pm$0.11 \\
74  & J140336.0+541924 & 1.55 & {\bf 13.2$\pm$1.0} & {\bf 4.8$\pm$0.6}  & 0.8$\pm$0.3  & {\bf 3.0$\pm$0.3}  & {\bf 1.9$\pm$0.2}  & {\bf 0.6$\pm$0.1}  & 2.80$\pm$0.16  & 1.74$\pm$0.10  & -0.37$\pm$0.04 & -0.59$\pm$0.07 & S,L\\
75  & J140336.0+543248 & 1.94 & {\bf 5.8$\pm$1.0}  & 0.7$\pm$0.5  & 0.0$\pm$0.3  & {\bf 1.2$\pm$0.2}  & 0.6$\pm$0.2  & 0.1$\pm$0.1  & 0.83$\pm$0.16  & 0.51$\pm$0.10  & -0.58$\pm$0.11 & -0.82$\pm$0.32 \\

\hline
\end{tabular}
}
\begin{tabular}{l}
\end{tabular}
\label{table:srclist}
\end{table*}

\begin{table*}
\contcaption{}
 \centering
{\scriptsize
  \begin{tabular}{@{}lccccccccccccc@{}}
\hline

Src & XMMU         & r$_{1\sigma}$ & \multicolumn{3}{c}{PN count rate (count ks$^{-1}$)} & \multicolumn{3}{c}{MOS count rate (count ks$^{-1}$)} & F$_X$   & L$_X$      & HR1            & HR2   & Var \\
    &                  & ($^{\prime\prime}$)   & S      & M            & H            & S            & M            & H            &    &                   &          &     & \\
(1) & (2)              & (3)    & \multicolumn{3}{c}{(4)}              & \multicolumn{3}{c}{(5)}                    & (6)            & (7)            & (8)            & (9)  & (10) \\

\hline

76  & J140336.3+541815 & 1.88 & 0.0$\pm$0.2  & {\bf 1.6$\pm$0.4}  & {\bf 1.6$\pm$0.4}  & 0.0$\pm$0.1  & {\bf 0.5$\pm$0.1}  & {\bf 0.5$\pm$0.1}  & 1.00$\pm$0.14  & 0.62$\pm$0.09  & 0.91$\pm$0.18  & 0.04$\pm$0.12 \\
77  & J140341.2+541902 & 1.65 & {\bf 9.5$\pm$0.9}  & {\bf 2.8$\pm$0.5}  & 1.2$\pm$0.4  & {\bf 2.5$\pm$0.3}  & {\bf 0.8$\pm$0.1}  & 0.5$\pm$0.1  & 2.14$\pm$0.16  & 1.33$\pm$0.10  & -0.53$\pm$0.05 & -0.31$\pm$0.11 & S\\
78  & J140341.8+543138 & 1.75 & {\bf 2.8$\pm$0.7}  & {\bf 3.8$\pm$0.8}  & 1.9$\pm$0.8  & {\bf 0.5$\pm$0.2}  & {\bf 1.1$\pm$0.2}  & {\bf 1.2$\pm$0.2}  & 2.21$\pm$0.27  & 1.37$\pm$0.17  & 0.26$\pm$0.11  & -0.07$\pm$0.11 & L\\
79  & J140344.3+542808 & 1.88 & {\bf 4.3$\pm$0.7}  & 0.0$\pm$0.2  & 0.0$\pm$0.3  & {\bf 0.8$\pm$0.2}  & 0.1$\pm$0.1  & 0.1$\pm$0.1  & 0.52$\pm$0.13  & 0.32$\pm$0.08  & -0.95$\pm$0.08 & -0.14$\pm$0.70 \\
80  & J140344.3+542700 & 2.01 & {\bf 2.1$\pm$0.5}  & 1.1$\pm$0.4  & 0.8$\pm$0.5  & {\bf 0.5$\pm$0.1}  & 0.1$\pm$0.1  & 0.4$\pm$0.2  & 0.81$\pm$0.18  & 0.50$\pm$0.11  & -0.41$\pm$0.16 & 0.11$\pm$0.25 \\
81  & J140344.5+542403 & 1.79 & 1.0$\pm$0.4  & 0.9$\pm$0.3  & 1.3$\pm$0.4  & 0.2$\pm$0.1  & {\bf 0.8$\pm$0.2}  & {\bf 0.8$\pm$0.1}  & 1.32$\pm$0.16  & 0.82$\pm$0.10  & 0.33$\pm$0.15  & 0.07$\pm$0.12 \\
82  & J140345.9+541617 & 1.68 & 1.7$\pm$1.3  & {\bf 8.9$\pm$1.9}  & {\bf 9.0$\pm$2.1}  & 0.2$\pm$0.1  & {\bf 1.0$\pm$0.2}  & {\bf 1.1$\pm$0.2}  & 2.25$\pm$0.25  & 1.40$\pm$0.15  & 0.63$\pm$0.13  & 0.04$\pm$0.09 \\
83  & J140346.4+543204 & 2.03 & {\bf 4.9$\pm$1.0}  & {\bf 3.8$\pm$0.9}  & 1.8$\pm$0.8  & -            & -            & -            & 2.12$\pm$0.47  & 1.32$\pm$0.29  & -0.12$\pm$0.16 & -0.36$\pm$0.22 \\
84  & J140347.3+541704 & 1.67 & {\bf 5.2$\pm$0.7}  & {\bf 2.1$\pm$0.5}  & 1.2$\pm$0.4  & {\bf 1.1$\pm$0.2}  & {\bf 0.8$\pm$0.1}  & 0.3$\pm$0.1  & 1.36$\pm$0.16  & 0.85$\pm$0.10  & -0.29$\pm$0.08 & -0.38$\pm$0.14 \\
85  & J140350.3+542413 & 1.61 & {\bf 7.7$\pm$0.8}  & {\bf 3.1$\pm$0.5}  & {\bf 2.3$\pm$0.5}  & {\bf 1.3$\pm$0.2}  & {\bf 1.4$\pm$0.2}  & {\bf 0.6$\pm$0.1}  & 2.16$\pm$0.18  & 1.34$\pm$0.11  & -0.26$\pm$0.07 & -0.32$\pm$0.09 & L\\
86  & J140351.2+541850 & 1.65 & {\bf 2.1$\pm$0.5}  & {\bf 1.8$\pm$0.4}  & {\bf 1.6$\pm$0.5}  & {\bf 0.8$\pm$0.1}  & {\bf 0.6$\pm$0.1}  & {\bf 0.7$\pm$0.1}  & 1.50$\pm$0.17  & 0.93$\pm$0.10  & -0.11$\pm$0.10 & 0.03$\pm$0.12 \\
87  & J140353.7+542159 & 1.94 & {\bf 7.2$\pm$0.8}  & {\bf 1.5$\pm$0.4}  & 0.5$\pm$0.4  & {\bf 1.5$\pm$0.2}  & {\bf 0.7$\pm$0.1}  & 0.0$\pm$0.1  & 1.06$\pm$0.13  & 0.66$\pm$0.08  & -0.55$\pm$0.07 & -0.80$\pm$0.19 & L\\
88  & J140353.7+541559 & 1.75 & 0.5$\pm$0.3  & {\bf 1.9$\pm$0.4}  & {\bf 3.3$\pm$0.6}  & 0.0$\pm$0.1  & {\bf 0.8$\pm$0.2}  & {\bf 1.3$\pm$0.2}  & 2.13$\pm$0.23  & 1.32$\pm$0.14  & 0.81$\pm$0.16  & 0.24$\pm$0.09 \\
89  & J140353.9+540939 & 1.58 & {\bf 7.2$\pm$1.0}  & {\bf 9.4$\pm$1.1}  & {\bf 4.6$\pm$0.9}  & {\bf 3.4$\pm$0.5}  & {\bf 4.7$\pm$0.6}  & {\bf 2.4$\pm$0.4}  & 5.66$\pm$0.42  & 3.51$\pm$0.26  & 0.15$\pm$0.06  & -0.34$\pm$0.07 \\
90  & J140354.0+540853 & 1.86 & {\bf 3.4$\pm$0.9}  & 2.2$\pm$0.7  & 0.5$\pm$0.6  & {\bf 1.3$\pm$0.3}  & {\bf 1.5$\pm$0.3}  & 0.1$\pm$0.2  & 1.24$\pm$0.25  & 0.77$\pm$0.15  & -0.02$\pm$0.11 & -0.76$\pm$0.20 & S\\
91  & J140357.0+541421 & 1.77 & {\bf 12.3$\pm$2.0} & 0.9$\pm$0.8  & 2.0$\pm$1.2  & -            & -            & -            & 2.59$\pm$0.71  & 1.61$\pm$0.44  & -0.86$\pm$0.11 & 0.36$\pm$0.46 & L\\
92  & J140359.7+540913 & 1.65 & {\bf 7.2$\pm$1.2}  & {\bf 5.7$\pm$1.0}  & {\bf 4.7$\pm$1.0}  & {\bf 2.1$\pm$0.4}  & {\bf 3.5$\pm$0.5}  & {\bf 2.1$\pm$0.4}  & 4.82$\pm$0.43  & 2.99$\pm$0.27  & 0.10$\pm$0.08  & -0.19$\pm$0.09 & L\\
93  & J140400.4+542533 & 2.02 & 1.3$\pm$0.5  & 1.4$\pm$0.5  & 1.0$\pm$0.5  & 0.2$\pm$0.1  & {\bf 0.7$\pm$0.1}  & 0.1$\pm$0.1  & 0.68$\pm$0.16  & 0.42$\pm$0.10  & 0.28$\pm$0.17  & -0.42$\pm$0.20 \\
94  & J140400.9+541132 & 1.94 & {\bf 6.4$\pm$1.6}  & 2.6$\pm$1.2  & 0.0$\pm$1.1  & {\bf 1.6$\pm$0.3}  & 1.0$\pm$0.3  & 0.5$\pm$0.3  & 1.64$\pm$0.37  & 1.02$\pm$0.23  & -0.30$\pm$0.12 & -0.37$\pm$0.27 \\
95  & J140402.9+540831 & 2.51 & {\bf 3.9$\pm$1.0}  & 0.7$\pm$0.6  & 0.9$\pm$0.7  & -            & -            & -            & 1.05$\pm$0.44  & 0.65$\pm$0.27  & -0.70$\pm$0.23 & 0.15$\pm$0.57 \\
96  & J140405.7+541602 & 1.74 & 0.0$\pm$0.1  & {\bf 2.2$\pm$0.5}  & {\bf 5.9$\pm$0.9}  & 0.2$\pm$0.1  & 0.6$\pm$0.2  & {\bf 1.9$\pm$0.3}  & 3.25$\pm$0.30  & 2.02$\pm$0.19  & 0.98$\pm$0.07  & 0.48$\pm$0.08 \\
97  & J140406.5+542310 & 2.03 & 1.6$\pm$0.5  & {\bf 2.0$\pm$0.5}  & 1.8$\pm$0.6  & {\bf 0.8$\pm$0.1}  & 0.3$\pm$0.1  & 0.1$\pm$0.1  & 0.90$\pm$0.18  & 0.56$\pm$0.11  & -0.17$\pm$0.15 & -0.10$\pm$0.19 \\
98  & J140406.7+541223 & 1.77 & {\bf 3.8$\pm$0.8}  & {\bf 2.5$\pm$0.7}  & 1.6$\pm$0.6  & {\bf 1.3$\pm$0.2}  & {\bf 1.2$\pm$0.3}  & 0.7$\pm$0.3  & 1.86$\pm$0.27  & 1.16$\pm$0.17  & -0.11$\pm$0.11 & -0.24$\pm$0.15 \\
99  & J140408.1+542347 & 1.75 & {\bf 4.5$\pm$0.7}  & 1.5$\pm$0.5  & 1.7$\pm$0.6  & {\bf 1.3$\pm$0.2}  & {\bf 0.7$\pm$0.2}  & 0.8$\pm$0.2  & 1.82$\pm$0.24  & 1.13$\pm$0.15  & -0.40$\pm$0.10 & 0.05$\pm$0.15 \\
100 & J140411.3+542521 & 2.05 & {\bf 3.6$\pm$0.7}  & 0.6$\pm$0.5  & 0.1$\pm$0.3  & {\bf 0.8$\pm$0.2}  & 0.0$\pm$0.1  & 0.0$\pm$0.1  & 0.42$\pm$0.11  & 0.26$\pm$0.07  & -0.89$\pm$0.11 & -0.65$\pm$0.71 & L\\
101 & J140411.4+541724 & 1.86 & {\bf 4.4$\pm$0.7}  & 1.1$\pm$0.4  & 1.2$\pm$0.5  & {\bf 1.1$\pm$0.3}  & 0.9$\pm$0.3  & 0.6$\pm$0.3  & 1.43$\pm$0.25  & 0.88$\pm$0.16  & -0.43$\pm$0.11 & -0.09$\pm$0.20 \\
102 & J140414.1+542604 & 1.51 & {\bf 63.8$\pm$2.8} & {\bf 40.8$\pm$2.3} & {\bf 12.3$\pm$1.4} & {\bf 15.6$\pm$0.8} & {\bf 14.7$\pm$0.8} & {\bf 4.9$\pm$0.5}  & 20.10$\pm$0.61 & 12.47$\pm$0.38 & -0.14$\pm$0.03 & -0.52$\pm$0.03 & S,L\\
103 & J140416.2+541020 & 2.48 & 2.2$\pm$0.8  & {\bf 2.8$\pm$0.8}  & 1.4$\pm$0.8  & -            & -            & -            & 1.49$\pm$0.48  & 0.92$\pm$0.30  & 0.14$\pm$0.23  & -0.32$\pm$0.28 \\
104 & J140416.7+541614 & 1.56 & {\bf 12.7$\pm$1.2} & {\bf 11.5$\pm$1.1} & {\bf 7.3$\pm$1.0}  & {\bf 4.9$\pm$0.5}  & {\bf 4.4$\pm$0.5}  & {\bf 3.5$\pm$0.5}  & 7.98$\pm$0.47  & 4.95$\pm$0.29  & -0.05$\pm$0.05 & -0.17$\pm$0.06 & S,L\\
105 & J140421.7+541921 & 1.57 & {\bf 25.0$\pm$1.7} & {\bf 5.8$\pm$0.8}  & 0.8$\pm$0.5  & {\bf 5.0$\pm$0.5}  & {\bf 2.4$\pm$0.4}  & 0.5$\pm$0.2  & 4.01$\pm$0.26  & 2.49$\pm$0.16  & -0.54$\pm$0.04 & -0.71$\pm$0.09 & S,L\\
106 & J140425.0+542550 & 2.08 & {\bf 2.9$\pm$0.8}  & {\bf 2.6$\pm$0.7}  & 1.3$\pm$0.7  & -            & -            & -            & 1.44$\pm$0.42  & 0.90$\pm$0.26  & -0.06$\pm$0.19 & -0.32$\pm$0.27 \\
107 & J140425.8+542047 & 4.73 & -            & -            & -            & 0.1$\pm$0.1  & 0.4$\pm$0.2  & {\bf 1.0$\pm$0.2}  & 1.62$\pm$0.33  & 1.00$\pm$0.21  & 0.50$\pm$0.37  & 0.41$\pm$0.22 \\
108 & J140429.1+542353 & 1.62 & {\bf 23.3$\pm$1.8} & 2.1$\pm$0.5  & 0.0$\pm$0.3  & {\bf 5.9$\pm$0.6}  & {\bf 1.4$\pm$0.3}  & 0.1$\pm$0.1  & 3.05$\pm$0.22  & 1.89$\pm$0.13  & -0.78$\pm$0.03 & -0.92$\pm$0.16 & L\\

\hline
\end{tabular}
}
\begin{tabular}{@{}l@{}}
(1) source number; (2) XMMU source designation (J2000 coordinates); (3) 1$\sigma$ error radius (including a 1.5 arcsecond systematic error);\\
(4 \& 5) source count rates in soft (0.3--1\,keV), medium (1--2\,keV) \& hard (2--6\,keV) bands for the PN and MOS cameras, with the\\ significant source detections ($>4\sigma$) highlighted in bold; (6) source flux in units of $10^{-14} \ergcms$ in the broad (0.3--6\,keV) band; \\(7) source luminosity in units of $10^{38} \ergsec$ in the 0.3--6\,keV band (assuming a distance to M101 of 7.2\,Mpc); (8 \& 9) soft (HR1) and \\hard (HR2) hardness ratios (as defined in the text); (10) X-ray variable on short (S) and long (L) time-scales.\\
\end{tabular}
\end{table*}

\begin{table*}
\caption{M101 source cross-identifications.}
 \centering
{\footnotesize
  \begin{tabular}{lcccl}
\hline

Src & \rosat ID$^a$  & \multicolumn{2}{c}{\chandra ID} & Other \\
    & HRI(H)/PSPC(P) & Pence$^b$  & Kilgard$^c$        &        \\
\hline

1       & P3                      & -                     & -                  & [PMC2001] RX J140134.94+542029.2 Galaxy\\
2       & -                       & -                     & -                  &        \\
3       & -                       & -                     & -                  & GSC1251 Star\\
4       & -                       & -                     & -                  &         \\
5       & H3/P4                   & -                     & J140203.6+541830   & XMM-8        \\
6       & -                       & -                     & -                  &         \\ 
7       & -                       & -                     & J140206.8+542534   &         \\ 
8       & H4                      & -                     & -                  &         \\ 
9       & H5                      & -                     & -                  &         \\ 
10      & H6/P5                   & -                     & -                  & [WIP] AGN \\ 
11      & -                       & -                     & -                  &         \\ 
12      & -                       & -                     & -                  &         \\ 
13      & -                       & -                     & -                  &         \\ 
14      & H8                      & -                     & J140214.1+542310   &         \\ 
15      & -                       & -                     & J140214.8+542804   &         \\
        &                         & -                     & [J140217.0+542803] &        \\ 
16      & -                       & -                     & J140218.9+541833   &         \\ 
17      & H9/P6                   & -                     & J140222.2+541756   & XMM-6   \\ 
18      & -                       & -                     & -                  &         \\ 
19      & -                       & -                     & -                  &         \\ 
20      & H10/P8                  & -                     & J140228.3+541626   & XMM-4, NGC 5447, [TFR] 2,[HK83,HGGK] HII \\
21      & H11/P7                  & -                     & -                  & [WIP] Star \\
22      & H12                     & -                     & J140228.7+542632   &         \\
23      & -                       & -                     & -                  &         \\
24      & -                       & -                     & -                  &         \\
25      & H13/P9                  & -                     & J140229.9+542119   & GSC1275 Star \\
26      & H14                     & -                     & J140232.5+542002   &         \\
27      & -                       & -                     & -                  &         \\
28      & -                       & -                     & -                  &         \\
29      & -                       & -                     & -                  &         \\
30      & -                       & -                     & J140238.9+542344   &         \\
31      & -                       & -                     & -                  &         \\
32      & -                       & -                     & J140245.2+543221   &         \\
33      & H16                     & -                     & J140246.4+542152   &         \\
34      & H17/P10                 & -                     & J140247.0+542656   & [WIP] Galaxy \\
35      & -                       & -                     & -                  &         \\
36      & -                       & -                     & -                  &         \\
37      & -                       & -                     & J140249.1+542241   &         \\
38      & -                       & -                     & -                  &         \\
39      & -                       & -                     & -                  &         \\
40      & -                       & -                     & J140251.5+541748   & [MF] 22 SNR, [HGGK] HII \\
41      & H18/P11                 & 5                     & J140252.9+542112   & XMM-12  \\
42      & -                       & -                     & J140252.9+542719   &         \\
43      & -                       & 13                    & J140301.2+542342   &         \\
44      & P12                     & -                     & -                  & NGC 5455, [HK83] HII, SN1970G \\
45      & -                       & 17                    & J140303.9+542133   &         \\
46      & H19/P13                 & -                     & -                  & XMM-1, [MF] 37 SNR, [TFR] 5  \\
47      & -                       & 19                    & J140305.2+541753   &         \\
48      & H21/P14                 & -                     & J140306.1+540713   &         \\
49      & -                       & 25                    & J140307.4+541938   &         \\
        &                         & [21]                  & [J140306.0+541945] &        \\
50      & -                       & 29                    & J140308.4+542459   &         \\
51      & H23/P16                 & 40                    & J140312.5+542057   & XMM-10 (nucleus), [CC2002] HII\\
        &                         & 38                    & J140312.5+542053   &         \\

52      & -                       & -                     & -                  &         \\
53      & H24                     & 45                    & J140313.6+542010   & [CC2002] HII  \\
54      & -                       & 47                    & J140313.7+541909   & [HGGK,H69] HII  \\
        &                         & [42]                  & [J140312.8+541901] & [MF] 46 SNR (= P42) \\
55      & H25/P17 [H22]           & 51                    & J140314.3+541807   & XMM-2, [H69] HII  \\
        &                         & [48]                  & [J140313.9+541811] &     \\
56      & -                       & -                     & -                  & GSC0731 Star \\ 
57      & -                       & 57                    & J140315.8+541749   & [H69] HII  \\
58      & -                       & -                     & -                  & [HK83,HGGK] HII \\

\hline
\end{tabular}
}
\begin{tabular}{l}
\end{tabular}
\label{table:crossid}
\end{table*}

\begin{table*}
\contcaption{}
 \centering
{\footnotesize
  \begin{tabular}{lcccl}
\hline

Src & \rosat ID$^a$  & \multicolumn{2}{c}{\chandra ID} & Other \\
    & HRI(H)/PSPC(P) & Pence$^b$  & Kilgard$^c$        &        \\

\hline

59      & H26                     & 63                    & J140318.1+541823   &         \\ 
        &                         & [60]                  & [J140316.8+541835] &         \\
        &                         & [62]                  & [J140317.7+541836] &         \\
60      & H27                     & 64                    & J140318.7+542430   &         \\
61      & -                       & 66                    & J140320.3+541633   &         \\
62      & -                       & -                     & -                  &         \\
63      & H29/P19 [H30]           & 70                    & J140321.5+541946   & XMM-9, [HGGK] HII  \\
        &                         & [67]                  & -                  & [MF] 54 SNR (= P67)        \\
64      & -                       & 71                    & J140321.7+541920   &         \\
        &                         & [68]                  & [J140321.2+541908] &         \\
65      & H30/P19 [H29]           & 76                    & J140324.2+541949   & XMM-14, [HK83,HGGK] HII \\
66      & -                       & -                     & J140325.1+543026   & [IGA75] 22 Radio source  \\
67      & -                       & 85                    & J140327.1+541832   & [MF] 65 SNR, [HK83] HII   \\
68      & -                       & -                     & J140328.3+542858   &         \\
69      & -                       & 92                    & J140329.9+542058   &         \\
        &                         & [90]                  & [J140328.9+542059] &         \\
70      & -                       & 93                    & J140330.0+542229   & [BKS96] Star cluster  \\
        &                         & [94]                  & [J140330.7+542222] &         \\
71      & H33                     & 99                    & J140333.3+541760   & [CC2002] HII  \\
72      & -                       & -                     & -                  &         \\
73      & H35                     & 103                   & J140335.5+541709   &   \\
74      & H36/P21                 & 104                   & J140336.0+541925   & XMM-11, [MF] 83 SNR, [LCC2001] Stars in MF83  \\
75      & -                       & -                     & -                  &         \\
76      & -                       & 105                   & J140336.3+541816   &         \\
77      & H37/P22                 & 107                   & J140341.3+541904   & XMM-13, NGC~5461, [HK83,HGGK] HII\\
78      & P23                     & -                     & J140341.1+543138   & [WIP] Star   \\
79      & -                       & -                     & J140344.9+542809   &         \\
80      & -                       & -                     & J140344.7+542658   &         \\
81      & -                       & -                     & J140344.9+542401   &         \\
82      & -                       & -                     & J140346.2+541615   &         \\
83      & -                       & -                     & J140346.5+543202   &         \\
84      & -                       & -                     & J140347.5+541704   &         \\
85      & H38/P24                 & -                     & J140350.6+542413   & [WIP] AGN \\
86      & -                       & -                     & J140351.3+541848   &         \\
87      & H40/P25 [H39,H41]       & 110                   & J140354.0+542157   & NGC~5462, [HK83,CasHII] HII, [ECB2002] $\theta$ Radio source\\
88      & -                       & -                     & J140353.6+541558   &         \\
89      & -                       & -                     & J140354.2+540938   & [TFR] 6  \\
90      & -                       & -                     & J140354.1+540852   &         \\
91      & -                       & -                     & J140357.5+541416   &         \\
92      & H43/P26                 & -                     & J140359.3+540912   & [WIP] AGN  \\
93      & -                       & -                     & J140400.5+542533   &         \\
94      & H44/P27                 & -                     & J140400.8+541130   & [WIP] AGN \\
95      & -                       & -                     & -                  &         \\
96      & -                       & -                     & J140405.9+541602   &         \\
97      & -                       & -                     & -                  &         \\
98      & -                       & -                     & J140406.9+541223   & [TFR] 7   \\
99      & -                       & -                     & J140408.4+542347   &         \\
100     & -                       & -                     & -                  & [S71] HII \\
101     & -                       & -                     & J140411.8+541726   &         \\
102     & H45/P28                 & -                     & J140414.3+542604   & XMM-3, [TFR] 8  \\
103     & -                       & -                     & -                  &         \\
104     & H47/P30                 & -                     & J140416.8+541615   & XMM-5, [PMC2001] RX J140416.61+541618.2 BLAGN  \\
105     & H48/P31                 & -                     & J140421.9+541920   & GSC1069, [AG79] 10 Star  \\
106     & -                       & -                     & J140425.3+542555   &         \\
107     & -                       & -                     & -                  &         \\
108     & H49/P32                 & -                     & J140429.1+542353   & XMM-7, NGC5471, [CCG2002] NGC~5471B \\

\hline
\end{tabular}
}
\begin{tabular}{l}
Notes: ~$^a$\rosat ID numbers correspond to HRI (H) and PSPC (P) detections \citep{wangetal99}. ~$^b$\chandra ID numbers from \citet{pence01}\\

$^c$\chandra ID numbers from \citet{kilgard04}. For (b) \& (c), square brackets denote additional {\it Chandra} sources matched to within 15 arcseconds \\of the {\it XMM} positions. `NGC' denotes giant HII regions e.g. \citet{williams95}. Catalogue references: [GSC] Guide Star Catalogue;\\ 

[AG79] \citet{allen79}; [CasHII] \citet{sanduleak87}; [BKS96] \citet*{bresolin96}; [CC2002] \citet{cedres02}; \\

 [CCG2002] \citet{chen02}; [ECB2002] \citet*{eck02}; [H69] \citet{hodge69}; [HGGK] \citet{hodge90};\\

[HK83] \citet{hodge83}; [IGA75] \citet*{israel75}; [LCC2001] \citet{lai01}; [MF] \citet{matonick97};\\

[PMC2001] \citet*{page01}; [S71] \citet{searle71}; [TFR] \citet{trinchieri90}; [WIP] \citet{wangetal99}.\\

\end{tabular}
\end{table*}

\section{Source detection techniques}
\label{sec:src_detection}

We have performed multi-band source detection on EPIC images in the following energy bands; soft (0.3--1\,keV), medium (1--2\,keV) and hard (2--6\,keV). Note that we have chosen to exclude data above 6\,keV due to the relative domination of background components at these high energies. From the filtered images described in the previous section, exposure maps were created for each band using the {\small SAS} task {\small EEXPMAP}, and detector masks were created for each camera with {\small EMASK} to define the areas of the images suitable for source detection. 

The key stages of source detection are the recognition of sources and their subsequent parameterization. The recognition stage can be performed using either a box or a wavelet detection algorithm. The standard {\small SAS} source searching routine implements the sliding box algorithm {\small EBOXDETECT}, which simultaneously searches the three energy band images for significant sources using a $5\times5$ pixel window with a local background determined in a frame around the search box. However, for this study we have also chosen to implement the {\small SAS} source detection task {\small EWAVELET}, which searches for sources in single energy bands by convolving the input images with a circularly symmetric Mexican Hat wavelet function. This method is generally more efficient at separating close sources in confused areas such as the central region of M101, where the situation is further complicated by the presence of a substantial diffuse component.

Two sets of preliminary source lists were created for the PN and MOS using the two different methods (box and wavelet) as the initial source recognition stage, using first-order background maps created with {\small ESPLINEMAP}. Both sets of sourcelists were then parameterized with {\small EMLDETECT}, which performs maximum likelihood instrumental point spread function (PSF) fits to the source count distributions. In the case of the sliding box detection method, this was performed simultaneously in the three energy bands, whereas the PSFs were fitted separately for each band with the wavelet method. All sources were treated as point sources, i.e. no source extent was fitted. Since the background is high in the EPIC instruments and the PSF wings extensive, we chose to improve the modeling of the background using the {\small SAS} task {\small ASMOOTH}, which adaptively smoothes the source-subtracted background images to a signal-to-noise ratio of 30. We ran {\small ASMOOTH} and {\small EMLDETECT} two times each on both sets of sourcelists, with the aim of improving the source parameterization with each iteration. The source detection threshold was set using a limiting likelihood (DET\_ML) of 11, which corresponds to a detection probability of 4$\sigma$ for a three-band detection. However, sources were only deemed to be significant if they reached a 4$\sigma$ single-band detection threshold (DET\_ML=10) in at least one of the three energy bands, and all sources not fulfilling this criterion were rejected. 

The two sets of sourcelists were then compared, and any additional sources detected with {\small EWAVELET} were added to the {\small EBOXDETECT} input list. The source parameterization process was then repeated to ensure a consistent multi-band parameterization for all sources detected by both methods. At a 4$\sigma$ detection threshold, we expect $\sim$ 6 spurious sources in the combined EPIC sourcelist due to statistical background fluctuations (i.e. $\sim$ 1 for each of the three energy band for both PN and MOS). To eliminate these, each source was visually inspected simultaneously in each detector in each energy band, and those which were deemed to be spurious (or at chip edges) were removed. In addition, sources falling outside of the \d25 ellipse of M101 were rejected. The source count rates and fluxes were determined by {\small EMLDETECT}, using the exposure maps to correct for vignetting and losses due to chip gaps and bad pixels/columns. To convert the count rates to fluxes, we computed energy conversion factors (ECFs) in {\small WEBSPEC} for each energy band assuming a simple power-law spectral shape with $\Gamma$=1.7 and Galactic absorption ($N_H\sim1\times10^{20} \cms$). Finally, the PN and MOS sourcelists were merged into a summary sourcelist with the {\small SAS} task {\small SRCMATCH}. Two detections of a source were merged if their positions matched to within their 3$\sigma$ errors (comprised of both a statistical error plus a 1 arcsecond systematic error).

\begin{figure*}
\centering
\includegraphics[width=14cm, height=13.5cm]{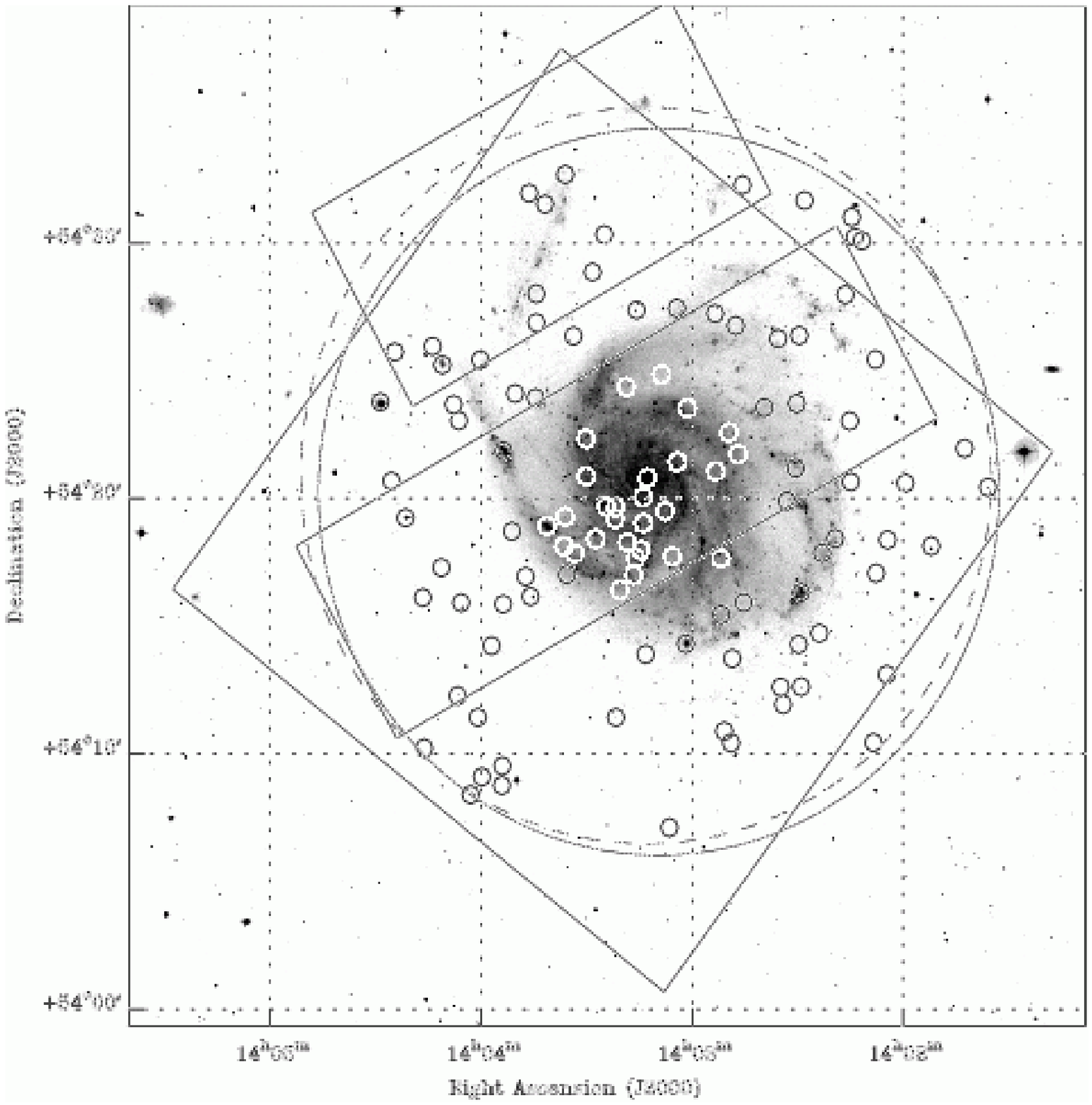}
\caption{\xmm source positions (circles) overlaid on a DSS2 optical blue image (note that source circles do not represent positional error radii). The large square and circular footprints denote the fields of view of the PN and MOS cameras respectively. The smaller rectangle footprints denote the fields of view of the ACIS-S and -I arrays in the 100\,ks \chandra observation of March 2000. The \d25 ellipse is shown with a dashed line.}
\label{fig:dssim}
\end{figure*}

\section{Source Catalogue}
\label{sec:catalogue}

In total, 108 sources were detected within the \d25 ellipse of M101 in the \xmm EPIC data. The source parameters are listed in Table~\ref{table:srclist} as follows: (1) source number, (2) XMMU designation, (3) 1$\sigma$ error radius including both the statistical error and a systematic error of 1.5 arcseconds added in quadrature, (4 \& 5) count rates in the soft (S), medium (M) and hard (H) bands in the PN and MOS cameras with significant band detections ($>4\sigma$) highlighted in bold, (6) flux in the 0.3--6\,keV band, (7) luminosity in the 0.3--6\,keV band assuming a distance to M101 of 7.2\,Mpc, (8 \& 9) soft (HR1) and hard (HR2) hardness ratios, (10) X-ray variability detected in the short- (S) and long- (L) term. 

The broad-band (0.3--6\,keV) fluxes quoted in column 6 were obtained by summing the fluxes over the three sub-bands and then taking the weighted mean of the individual PN and MOS measurements. The faintest source detected has a flux of 3.1$\times10^{-15} \ergcms$ in the 0.3--6\,keV range, and the brightest source (\#46/XMM-1) has a flux of 4.2$\times10^{-13} \ergcms$. These correspond to a luminosity range of 1.9$\times10^{37}$--2.6$\times10^{39} \ergsec$ at the distance of M101. In Figure~\ref{fig:xmm}, the full set of sources is plotted on a broad-band (0.3--6\,keV) image EPIC of M101 (a), and separate images are shown for the soft (b), medium (c) and hard bands (d) with significant ($>4\sigma$) detections in those bands marked with squares (PN) and triangles (MOS). To illustrate the source positions with respect to the optical emission from M101, Figure~\ref{fig:dssim} shows the source positions overlaid on a DSS2 blue image.

The hardness ratios shown in columns 8 \& 9 are calculated directly from the source count rates measured in the source detection routines. They are defined as HR1=(M-S)/(M+S) and HR2=(H-M)/(H+M) where S, M \& H denote the count rates in the three energy bands. In order to take advantage of the improved statistics, we quote the weighted mean EPIC (combined PN and MOS) ratios for sources detected in both cameras.  For sources detected only in either the PN or MOS data, the single camera ratios are listed (see section~\ref{sec:xcolours} for further details).

\begin{table*}
\caption{X-ray colour classifications for \xmm EPIC observations in the 0.3--6\,keV range (medium filter).}
 \centering
  \begin{tabular}{lccc}
\hline
Classification            & \multicolumn{3}{c}{Definition}  \\
                          & EPIC (PN \& MOS)                  & PN only                                  & MOS only                                 \\
\hline
Supernova remnant         & HR1$<$-0.24, HR2$<$-0.10             & HR1$<$-0.34, HR2$<$-0.14             & HR1$<$-0.15, HR2$<$-0.07             \\ 

X-ray binary              & -0.24$<$HR1$<$0.57, -0.8$<$HR2$<$0.8 & -0.34$<$HR1$<$0.52, -0.8$<$HR2$<$0.8 & -0.15$<$HR1$<$0.62, -0.8$<$HR2$<$0.8 \\

Background source         & HR1$<$-0.24, HR2$>$-0.10             & HR1$<$-0.34, HR2$>$-0.14             & HR1$<$-0.15, HR2$>$-0.07             \\

Absorbed source           & HR1$>$0.57                          & HR1$>$0.52                           & HR1$>$0.62                           \\

Indeterminate soft source & -0.24$<$HR1$<$0.57, HR2$<$-0.8       & -0.34$<$HR1$<$0.52, HR2$<$-0.8       & -0.15$<$HR1$<$0.62, HR2$<$-0.8       \\

Indeterminate hard source & -0.24$<$HR1$<$0.57, HR2$>$0.8        & -0.34$<$HR1$<$0.52, HR2$>$0.8        & -0.15$<$HR1$<$0.62, HR2$>$0.8        \\

\hline
\end{tabular}
\label{table:xcolours}
\end{table*}

\subsection{X-ray/multiwavelength cross-correlations}
\label{sec:ID}

We have cross-correlated the \xmm sourcelist with previous X-ray observations of M101 (see Table~\ref{table:crossid}). For the \chandra observations detailed in section~\ref{sec:obs}, we matched on-axis sources (whose positions are generally accurate to $\sim$ 1 arcsecond) to within the \xmm 3$\sigma$ errors. For off-axis sources, the decreasing \chandra positional accuracy to $\sim$ 2 arcseconds was also taken into account. However, given the large PSF of \xmm ($\sim$ 6 arcseconds FWHM), we have also checked for any contamination from additional fainter sources detected only by \chandra by searching for sources that lie within 15 arcseconds of the \xmm source positions (this corresponds to the on-axis 68 per cent energy cut-out radius used in {\small EMLDETECT}). In total, 71 \xmm sources are unambiguously matched to single \chandra sources within the 3$\sigma$ errors, whereas the nuclear source is resolved into two sources by {\it Chandra}. These matches are listed in Table~\ref{table:crossid}, and any additional sources matching to within 15 arcseconds are shown in square brackets. For completeness, we show both the CXOU designations of \citet{kilgard04} and equivalent source numbers from \citet{pence01}.

The \rosat HRI and PSPC X-ray detections of \citet{wangetal99} were matched to within the combined 3$\sigma$ \xmm and \rosat errors, and 38 matches were found. Where sources detected in the PSPC observation were confused with multiple HRI sources, the confused HRI sources are shown in square brackets. In addition, in the absence of any other multiwavelength identification, the optical counterpart identifications for several sources from \citet{wangetal99} are also included (denoted [WIP] in the final column of Table~\ref{table:crossid}). The \einstein source detections of \citet{trinchieri90} (denoted [TFR]) are also shown in the final column.

To aid our identification of the X-ray sources, we have searched for multi-wavelength correlations in the NED and SIMBAD archives using a 3$\sigma$ search radius, and the results are also shown in the final column in Table~\ref{table:crossid}. The XMM--n source designations of the bright sources in Paper I are also listed here. In addition, we have searched for matches to the 93 optical SNRs found in M101 by \cite{matonick97} (denoted [MF]), matching with the more accurate \chandra positions where available to within a combined 2 arcsecond error. Four matches were found; three using \chandra positions (\xmm source numbers \#40, 67 \& 74), and one with the bright \xmm source (\#46/XMM-1) whose position matches that of MF37 to $\sim$ 2 arcseconds. We also list the matches between the faint \chandra source P67 (confused with \#63/XMM-9) and MF54, whose positions correlate to within less than an arcsecond (as demonstrated by \citealt{snowden01}), plus the \chandra source J140312.8+541901 (confused with \xmm source \#54), which is coincident with MF46 to within $\sim$ 1 arcsecond.

Overall, 74 \xmm sources are X-ray detections only. Of the remainder, 20 are coincident with \hii and/or SNRs, 7 have identified/candidate background AGN/galaxy counterparts, 6 are coincident with foreground stars and one has a radio counterpart.

\section{X-ray Colours}
\label{sec:xcolours}

In Paper I, we performed detailed spectral analyses of the fourteen brightest sources in the \xmm field with sufficient counts for spectral fitting ($>$ 300 in the PN data). Here, we attempt to broadly classify the complete \xmm source population according to their X-ray colours (hardness ratios). Although we cannot definitively classify any source by its X-ray colour alone, this approach can be a useful starting point for source identification as well as giving insights into the overall source population of a galaxy.

\begin{figure*}
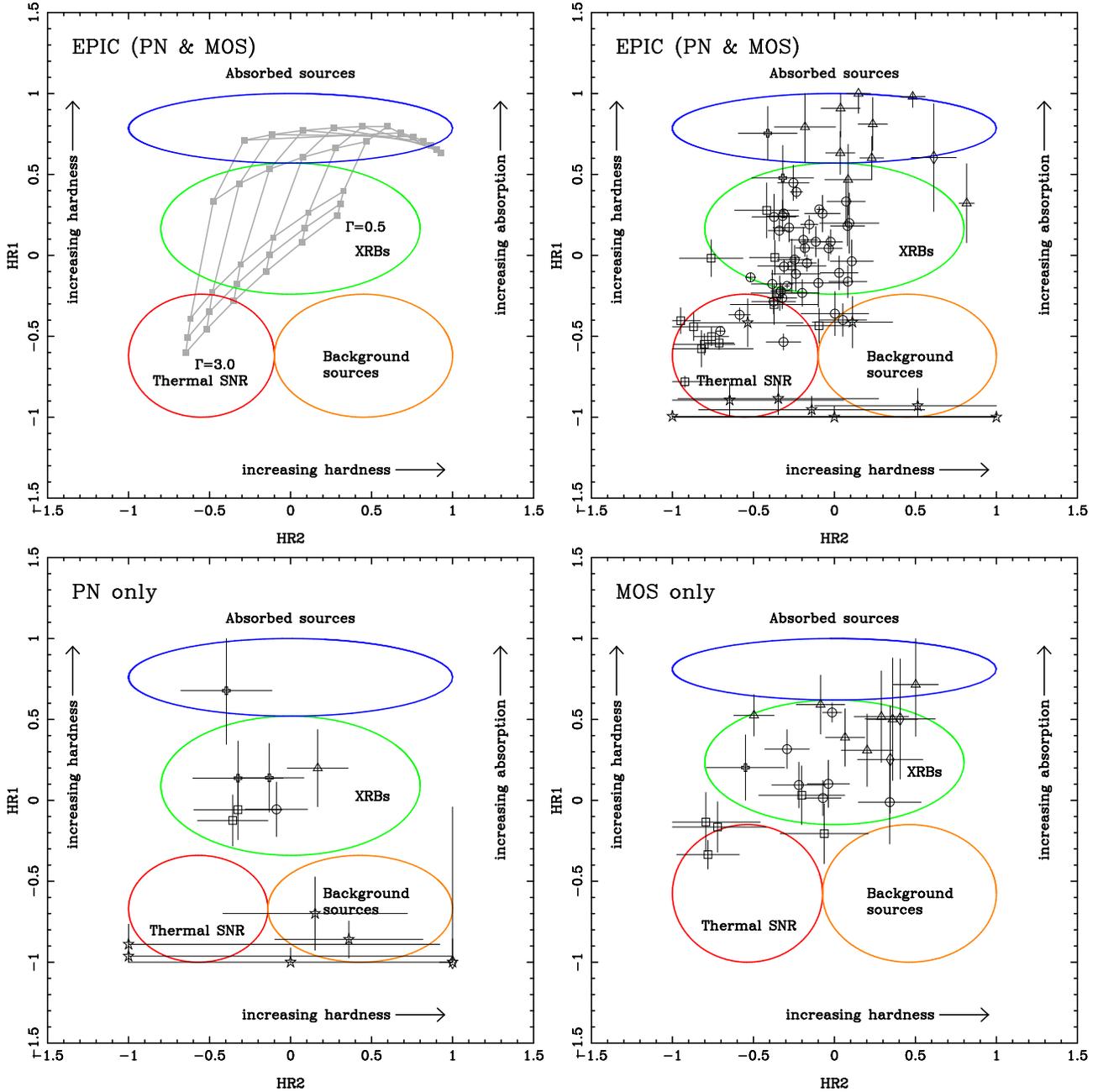

\centering
\rotatebox{270}{\scalebox{0.55}{\includegraphics{figure3a.ps}}}
\rotatebox{270}{\scalebox{0.55}{\includegraphics{figure3b.ps}}}
\rotatebox{270}{\scalebox{0.55}{\includegraphics{figure3c.ps}}}
\rotatebox{270}{\scalebox{0.55}{\includegraphics{figure3d.ps}}}
\caption{Upper left: \xmm EPIC (PN \& MOS) X-ray colour diagram showing the source classification scheme defined by Prestwich et al. (2003) and Kilgard et al. (2004). The ellipses illustrate the boundaries of the colour ranges for each category. The model grid shows $N_H$ absorption values ranging between $10^{20}-5\times10^{22} \cms$ (bottom to top) with photon indices ranging between $\Gamma=0.5-3.0$. Upper right: \xmm EPIC (PN \& MOS) X-ray colours of the discrete sources in M101 detected with both cameras. The symbols denote true (T) or false (F) detections above the 3$\sigma$ detection threshold in the soft, medium and hard bands: circle (TTT), box (TTF); triangle (FTT); star (TFF); cross (FTF); diamond (FFT). Lower left: \xmm PN only detections. Lower right: \xmm MOS only detections.}
\label{fig:xcolours}
\end{figure*}

We have used the X-ray colour classification scheme of \citet{prestwich03} and \citet{kilgard04}, which was developed to classify \chandra point sources using the colours of known XRBs and SNRs. We have modified the scheme for \xmm EPIC data, ensuring that the source categories cover the same spectral ranges as the \chandra scheme (Figure~\ref{fig:xcolours}). The distribution in this colour space of sources detected with both PN and MOS (i.e. `EPIC') with $\Gamma$=0.5--3 and absorption values of $N_H=10^{20}-5\times10^{22} \cms$ is shown in Figure~\ref{fig:xcolours} (top left), with ellipses illustrating the upper and lower boundaries of the coarse colour ranges for different source categories (see Table~\ref{table:xcolours} for details). For consistency with the data (see section~\ref{sec:catalogue}), the EPIC power-law grid was calculated using the mean of the separate model PN and MOS hardness ratios at each point. The XRB range encompasses sources with spectral shapes equivalent to a power-law slope of $\Gamma\sim$1--2.5, typical of low- and high-mass XRBs (\citealt{prestwich03} and references therein), and the SNR range covers soft sources ($\Gamma\ga2$) with low absorption, typical of the soft thermal spectra of known SNRs. The absorbed source range applies to sources with $N_H\ga5\times10^{21} \cms$, and the background source range covers objects with complex two-component spectra with highly absorbed power-law continua and soft excesses.

In Figure~\ref{fig:xcolours} (top right) the X-ray colours of the sources detected in both the PN and MOS cameras are plotted. For sources detected only in either the PN or MOS data, we have plotted their separate hardness ratios with equivalent classification spectral ranges applicable to the PN and MOS responses in Figure~\ref{fig:xcolours} (bottom).  In general, the sources detected by PN only are near MOS chip gaps/detector edges or are very soft sources as is shown in Figure~\ref{fig:xcolours} (bottom left). This higher sensitivity to soft sources is to be expected, as the PN camera has a much greater effective area at soft energies than the MOS cameras. For the MOS only sources (Figure~\ref{fig:xcolours}, bottom right), the distribution of hardness ratios is more even, and they were generally not detected in the PN data because they were either outside the PN field of view or near PN chip gaps.

Since the sources are expected to (and indeed have) a variety of spectral shapes, they are not all detected significantly in all three energy bands.  We have therefore plotted the sources in the X-ray colour plane using the following symbols: circle (TTT), box (TTF); triangle (FTT); star (TFF); cross (FTF); diamond (FFT) denoting sources detected (T) or not detected (F) significantly in the soft, medium and hard bands respectively. For this purpose, a 3$\sigma$ detection threshold is used (rather than the 4$\sigma$ used for source detection) to provide a better estimation of the intrinsic spectral shapes of the sources.

\begin{figure*}
\centering
\includegraphics[width=12cm, height=11.3cm]{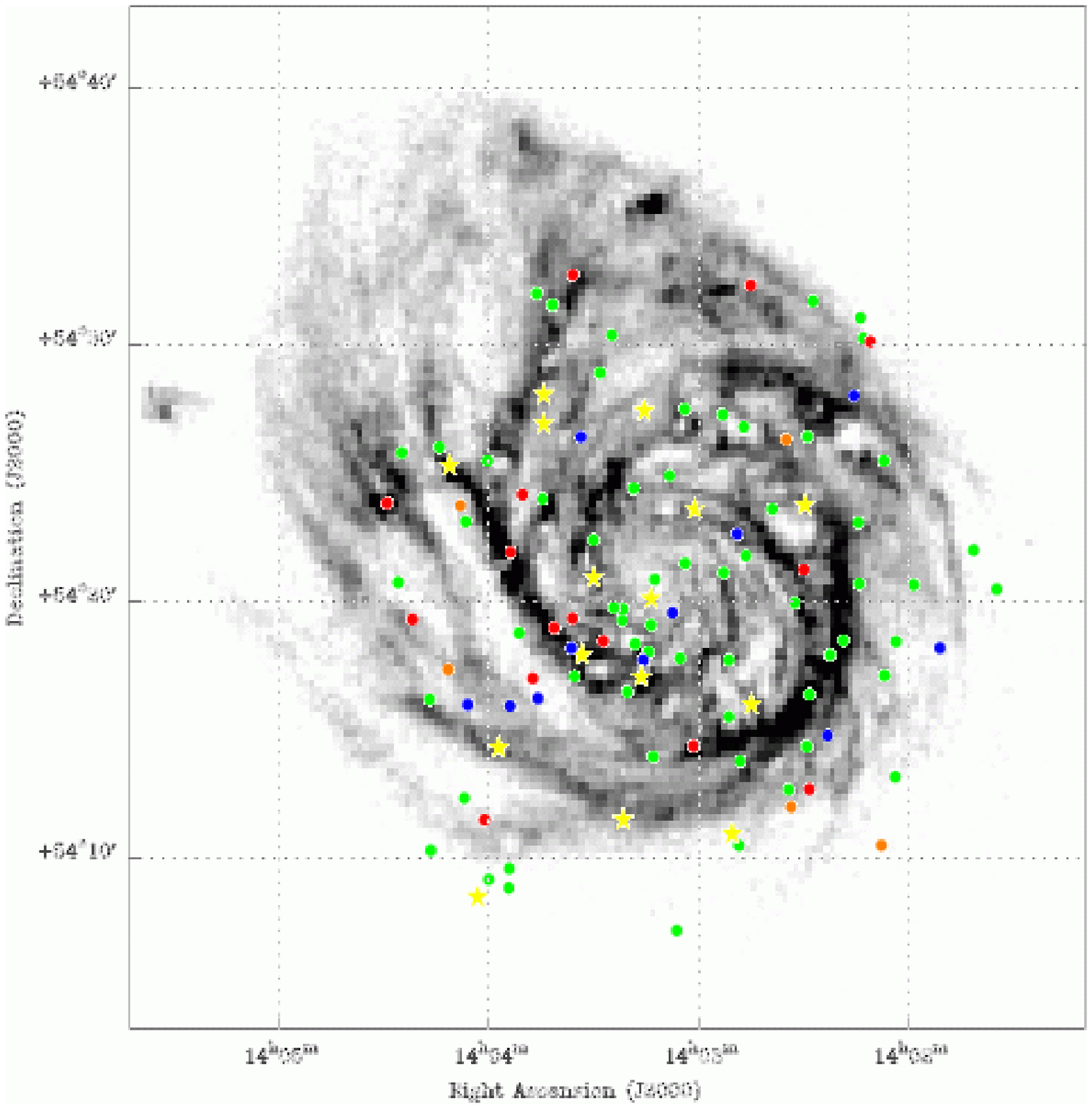}
\includegraphics[width=12cm, height=11.3cm]{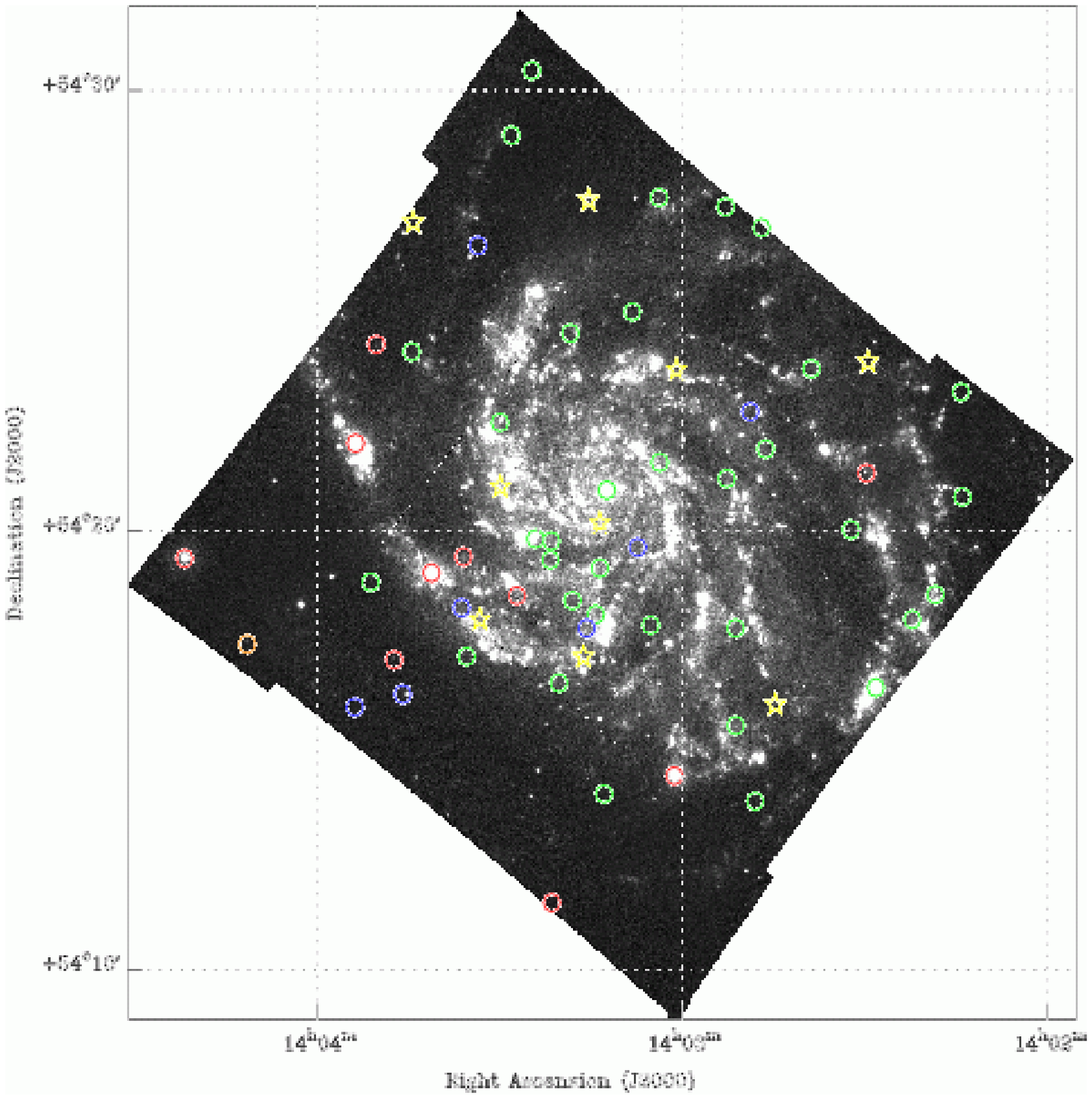}
\caption{Top: \xmm sources overlaid on an \hi image of M101 from Kamphuis, Sancisi, \& van der Hulst (1991).  Bottom: \xmm sources overlaid on an \xmm OM UVW1 ($\lambda_{\mathrm{central}}=2910$ \AA) image of the central region of M101. The colours corresponding to the source classifications in the X-ray colour scheme, and candidate SSSs are marked with yellow stars.}
\label{fig:im_nh}
\end{figure*}

The sources are distributed among the four designated colour-colour categories as follows: 62 XRBs, 15 SNRs, 11 absorbed sources and 5 background sources. In addition, there are 15 candidate SSSs i.e. sources only detected significantly in the soft band (see section~\ref{sec:sss}). These are not included in any of the four categories, but appear at HR1$\simeq$-1 (marked with stars in Figure~\ref{fig:xcolours}). However, additional information such as source variability, position in the galaxy and multiwavelength counterparts should be taken into account when classifying sources, and each category is discussed in turn in section~\ref{sec:class}. In Figure~\ref{fig:im_nh}, the positions of the sources are plotted on an \hi map and an \xmm Optical Monitor (OM) UV image of M101, with colours denoting where the sources lie in the classification scheme.

\section{Timing properties}
\label{sec:timing}

\subsection{Short-term variability}

In Paper I, eleven of the fourteen brightest sources in the \xmm observation were found to be variable at $>$ 95 per cent probability using $\chi^2$, Kolmogorov-Smirnov (K-S) and standard deviation tests (see table 5, Paper I).  For the majority of the remainder of the sources, we do not have sufficient counts for such detailed analyses. Instead, we have performed simple standard deviation tests in the following manner. Firstly, for consistency between the datasets and to filter out as much background variability as possible, the heavier time filter used in the PN source detection was applied to the MOS data. Sources were then divided into categories of 2, 4, 6 \& 8 time bin resolutions depending on their total (background subtracted) counts (PN+MOS) detected during this period ($\sim$ 26\,ks), ensuring at least 20 counts per bin for each source. Images were created for each time bin category using equally-spaced time intervals, and the source counts in each determined with {\small EMLDETECT}. 

\begin{figure}
\centering
\rotatebox{270}{\scalebox{0.48}{\includegraphics{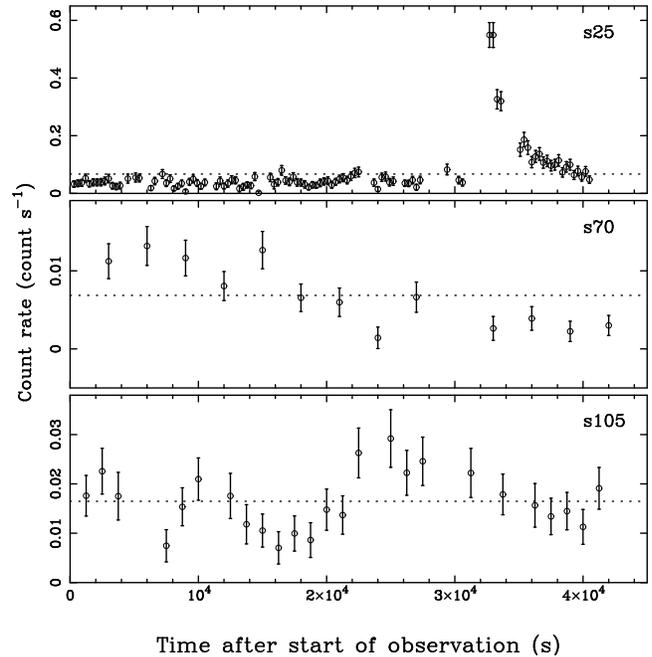}}}
\caption{\xmm short-term lightcurves for three variable sources (\#25, 70 \& 105), obtained by summing the counts from the 3 EPIC cameras. The mean count rates are shown as dashed lines and error bars correspond to 1$\sigma$ deviations assuming Gaussian statistics.}
\label{fig:lcurve}
\end{figure}

Fifty-one sources had sufficient counts for a 2 time-bin resolution only, and for these we compared the two values with their respective 1$\sigma$ errors. Of these, only two were variable above the 95 per cent level (\# 19 at 2.7$\sigma$ \& \# 90 at 2.8$\sigma$). For the remaining sources with 4, 6 \& 8 time bins, we calculated a standard deviation of the number of counts per bin from the mean, and compared this to the expected deviation of $\sim$ 22 per cent expected from Gaussian counting noise. Only three sources (\# 25, 70 \& 105) were variable at $>$ 95 per cent. As each had comparably high numbers of counts ($>$ 200), we confirmed their variability by extracting full short-term light-curves, again with time bin sizes tailored so so that each bin had at least 20 counts after background subtraction. Again, to minimize contamination from soft proton flaring, we applied the heavier time filtering to the data, and corrected the exposure times and count rates in incomplete bins with a simple scaling factor, excluding bins with $<$0.3 times the exposure remaining of the original bin size. The resulting light curves are shown in Figure~\ref{fig:lcurve}, and $\chi^2$ tests showed each to be variable at greater than 99.73 per cent probability.  Interestingly, two of these variable sources are the two bright foreground stars (\# 25 \& 105), which were excluded from our analyses in Paper I as they were not associated with M101 (see section~\ref{sec:stars} for further discussion). In total, 16 sources show short-term variability above the 95 per cent level in this dataset (including the variable bright sources from Paper I), which we denote with an `S' in column 10 of Table~\ref{table:srclist}.

\subsection{Long-term variability}

\subsubsection{Long-term flux changes}

To search for long-term source variability, we have compared the observed fluxes in the \xmm observation with those measured in the \chandra observations conducted $\sim$ 2 years previously, plus four archival \rosat HRI observations. The \rosat observations span $\sim$ 4 years beginning in January 1992, giving a total baseline of $\sim$ 10 years for sources detected with both \rosat and {\it XMM}. Twenty-three \xmm sources were detected in at least one of the four HRI observations, thirteen of which are the bright sources analysed in Paper I. Since the HRI count rates quoted in \citet{wangetal99} are averages of the four individual observations, we have re-analysed the archival data sets (as per \citealt{roberts00}), measuring count rates (or upper limits) in the 0.5--2\,keV band for each individual observation. Seventy one \xmm sources are matched in the \chandra observations, ten of which are confused (or possibly confused) with nearby sources. Where sources were detected in both \chandra observations, the source counts from the longer observation were used to ensure the most accurate flux determinations.  For consistency between the data sets, the measured count rates from the \chandra and \rosat observations were converted into fluxes in the 0.3--6\,keV band using ECFs determined from {\small WebPIMMS}, using the same spectral model used for the \xmm flux conversions ($\Gamma$=1.7, $N_H$=$1\times10^{20} \cms$). 

To quantify this variability, we have compared the maximum to minimum observed fluxes of each source in this multi-mission data set. For the {\it XMM}/\rosat sources shown to be confused by multiple source detections in the \chandra observations, flux changes were only deemed real in the \chandra data if the variability was only apparent using the flux of the source matching most closely to the \xmm position, as opposed to the combined flux of the confused sources. In total, 37 sources show long-term variability at $>2\sigma$, with 26 variable above the 3$\sigma$ level. The majority (32 of 37) vary by a factor of 1.5--4 between the highest and lowest observed fluxes. The most variable sources are \#21, 53 \& 87, which vary by factors of 6--8, and \#56 (a foreground star), which is $\sim$ 17 times fainter than in a \rosat observation in 1992. The most variable source we detect is \#55/XMM-2 (see section~\ref{sec:trans} below). 

In addition, 30 \xmm sources have not been detected in any previous X-ray observation, and we have investigated whether this is due to real source variability as opposed to differing coverage and/or lack of photon sensitivity. Fourteen sources are covered by the two \chandra observations; the flux limits on the positions of four (\#3, 13, 58 \& 100) imply that they have all brightened by at least factors of 3--5 between between the \chandra and \xmm observations, while eight are at or below the detection thresholds at their positions. The remaining two sources (\#24 \& 29) are sufficiently variable to be classified as transients (see section~\ref{sec:trans} below). The whole field was covered in the \rosat HRI observations, but all but one of the \xmm sources with no HRI matches fall below the flux limit of the longest 108\,ks HRI observation ($F_X\sim2.3\times10^{-14} \ergcms$, $L_X\sim1.4\times10^{38} \ergsec$, extrapolated to the 0.3--6\,keV band) and therefore cannot be classed as variable based on these data. The only exception is \#35, which should have been detected in the both of the longest HRI observations. The flux limits of these observations imply an increase in luminosity of this source of at least a factor of $\sim$ 2 in the \xmm observation.

In total, 44 \xmm sources ($\sim$ 40 per cent of the source population) show long-term variability, either by direct measurement with detections at different epochs (at $>2\sigma$) or by the implied variations of the flux limits of the archival data. These are all denoted by `L' in column 10 of Table\ref{table:srclist}.

\subsubsection{Transient sources}
\label{sec:trans}

Transient behaviour is characteristic of low-mass XRB (LMXB) systems, with observed outbursts believed to be caused by hydrogen-ionization instabilities in the accretion disc around the primary black hole/neutron star (see for example \citealt{king02a}). Such outbursts do not generally occur in high-mass XRBs (HMXBs), as the companion star is a sufficient source of irradiation to keep the accretion disc in a stable ionized state (although it is possible that they could occur in HMXBs with Be-star companions, see \citealt{king04}). Galactic soft X-ray transients (SXTs) have quiescent X-ray luminosities in the $10^{32}-10^{34} \ergsec$ range (see e.g. \citealt{menou99}), and often increase by a factor of $10^7$ or more in outburst \citep{king04}. Since such faint quiescent sources are beyond the flux limits of any existing X-ray observation of M101, we have searched for {\it candidate} transient sources by comparing the full \xmm sourcelist with previous \chandra and \rosat detections to ascertain lower limits on the flux variability of highly variable sources. For these purposes we define candidate transient sources as those varying by at least an order of magnitude.

The most obvious transient candidate detected in the \xmm observation is \#55/XMM-2, which has a luminosity a factor of $\sim$ 30 brighter than the flux level seen in previous observations (see Paper I). Another highly variable source is the most luminous ULX (P98) in the long \chandra observation \citep{mukai03}. It was initially detected in a \rosat observation (H32), and increased in flux by a factor of $\sim$ 5 in the later \chandra observation. It then fell below the detection threshold in the \xmm observation, implying another drop in luminosity of at least a factor of $\sim$ 20. For an additional two \xmm sources not detected by {\it Chandra}, the difference between their fluxes and the flux limits in the \chandra observations at those positions imply variability large enough to be classed as candidate transient behaviour. Source \#24 with SSS X-ray colours shows a massive increase of a factor of at least $\sim$ 580 in the \xmm observation, while \#29 must have increased by a factor of at least $\sim$ 20.

It is also worth noting that there are another four bright \rosat HRI sources ($L_X\ga10^{38} \ergsec$ in the 0.3--6\,keV band) that are undetected in the \xmm observation, suggesting transient behaviour. Two sources, H28 with a ULX luminosity of $L_X\sim10^{39} \ergsec$ in the first HRI observation, and H31 with $L_X\sim4\times10^{38} \ergsec$ in the second observation, are both undetected by \chandra and {\it XMM}. H15 is detected in the fourth HRI observation with $L_X\sim1.4\times10^{38} \ergsec$, but is then detected with \chandra with a factor of $\sim$ 5 lower flux.  Another source, H34, is detected in the combined HRI dataset by \citet{wangetal99}, with an average luminosity of $L_X\sim1\times10^{38} \ergsec$. Therefore, we identify eight candidate transient sources in total using these datasets.

\subsubsection{Colour changes}

Nineteen sources show variability at the 3$\sigma$ level between the \xmm and \chandra observations, ten increasing and nine decreasing in flux. Since XRBs and ULXs are known to undergo spectral transitions as their accretion rates change, we have compared the X-ray colours of these sources in both observations to search for such behaviour. To directly compare the source colours between the observations, correction factors were applied to the \chandra X-ray colours to account for the differences in energy response and effective area of the two telescopes. The correction factors were calculated as the differences in the soft and hard colours for representative average spectral models for XRBs ($\Gamma$=1.5, $N_H$=$10^{21} \cms$), SNRs ($\Gamma$=3, $N_H$=$10^{20} \cms$) and absorbed sources ($\Gamma$=1.5, $N_H$=$10^{22} \cms$), and were applied depending on the source positions on the equivalent \chandra colour-colour plane.

\begin{figure}
\centering
\rotatebox{270}{\scalebox{0.55}{\includegraphics{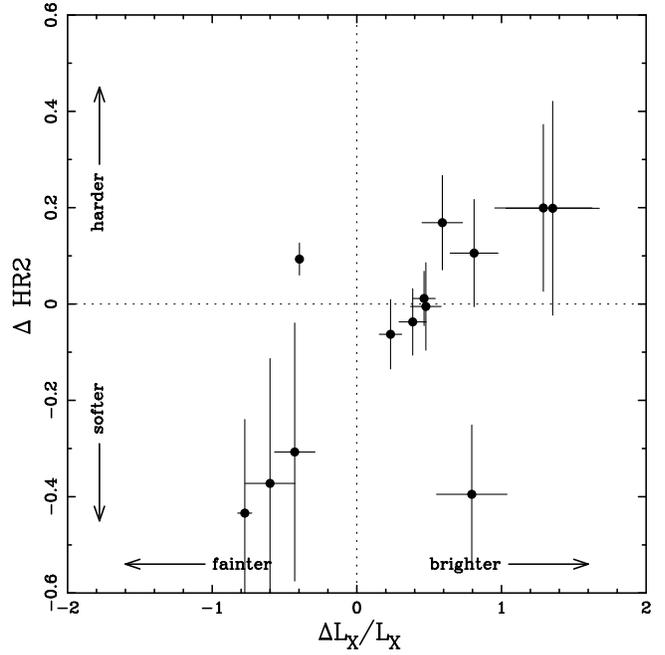}}}
\caption{Plot of the change in hard X-ray colour ($\Delta$HR2) versus the fractional change in luminosity [$\Delta L_X/L_X$=($L_{X(XMM)}-L_{X(Chan)})/L_{X(Chan)}$] between the \chandra and \xmm observations of M101.}
\label{fig:colour_changes}
\end{figure}

Figure~\ref{fig:colour_changes} shows the changes in the hard (HR2) X-ray colour versus the change in luminosity between the two observations as a fraction of the \chandra luminosity [$\Delta L_X/L_X$=($L_{X(XMM)}-L_{X(Chan)})/L_{X(Chan)}$]. Note that we have not used the soft (HR1) colour to look for intrinsic colour changes, as this is much more sensitive to varying absorption within M101. The two foreground stars are not plotted in this diagram, nor do we plot the SSSs as their hard colours are unconstrained. The transient source (\#55/XMM-2) is also excluded as its change in luminosity is far larger than the other sources ($\Delta L_X/L_X\sim$32) and it has no significant colour variation. Although these thirteen sources all show some shift in colour between the observations, all but two remain in the same spectral classification between the observations as expected. The exceptions are \#20/XMM-4, which increases in flux between the \chandra and \xmm observations, moving from the absorbed to the XRB class, and \#87 located in the giant \hii region NGC~5462, which decreases in flux and moves from the XRB to the SNR class (see section~\ref{sec:snr}).

Interestingly, the majority of sources do not appear to follow the `canonical' low/hard (LH) power-law dominated to high/soft (HS) disc-dominated state transitions seen in many Galactic black hole binary systems (e.g. \citealt{mcclintock04}). Only four sources follow this pattern using the HR2 colour (\#41/XMM-12, \#65/XMM-14, \#102/XMM-3 \& \#78), three of which are bright sources analysed in detail in Paper I. The remaining sources fall into the opposite quadrants where the fainter the source, the softer its spectrum and vice versa, and recent studies with \xmmn and \chandra have demonstrated that many bright ULXs do display this behaviour (e.g. \citealt{fabbiano03a}; \citealt{dewangan04}; \citealt{roberts04}). Although there may be underlying physical reasons for this type of behaviour in XRBs (see for example \citealt{fabbiano03a}), we must consider whether these results may partly be due to the X-ray energy bands covered by both \xmmn and \chandra in these studies (typically 0.3--10\,keV). Indeed, when using the broader band 2--8\,keV/0.3--2\,keV flux ratio in Paper I, seven sources did show a typical softening with increasing luminosity. The main spectral states of XRBs have traditionally been defined using a much harder 2--20\,keV energy range, covered by e.g. the Rossi X-ray Timing Explorer ({\it RXTE}, see \citealt{mcclintock04}). Since the sources in our sample have luminosities at $\ga10^{38} \ergsec$, they are possibly in disc-dominated states. If their inner disc temperatures are $\sim$1--2\,keV, they will contribute to the spectrum mainly at 1--6\,keV. If the flux change is coming from the emergence of the disc relative to the power-law component, the spectrum can appear to get harder in the 1--2\,keV range compared to the 2--6\,keV range. Therefore, we could still be observing the classic power-law to disc state transition; however observations of large samples of XRBs in this wider X-ray spectral band will be needed to confirm whether this is the case.

\section{Source classification}
\label{sec:class}

\subsection{Background galaxies/AGN}
\label{sec:agn}

We can estimate the probability of any of our detections being background sources using the hard-band (2--10\,keV) integral log($N$)--log($S$) relationship of \citet{campana01}. We judge this to be a better indicator than the soft-band log($N$)--log($S$) due to the high, spatially variable neutral absorption column associated with the disc of M101. A 4$\sigma$ source detection threshold in the hard 2--6\,keV band corresponds to $\sim$ 28 counts in both the PN  and MOS. After converting this into a 2--10\,keV flux for a typical faint AGN ($\Gamma$=1.4, $N_H=3\times10^{20} \cms$), and folding this through both the exposure map for the observation (to correct for sensitivity changes across the EPIC field-of-view) and the 2--10\,keV log($N$)--log($S$), we predict a total contamination of $\sim$ 18 background AGN in the PN data and $\sim$ 24 in the MOS.

Sources \#1 \& 104 have been optically identified as background galaxies in the \rosat survey of hard sources of \citet{page01}. We note that source \#104 is one of the bright sources analysed in Paper I (XMM-5), where we omitted this optical identification as a broad-line AGN. It showed a composite X-ray spectrum comprised of a power-law plus a {\small MEKAL} thermal plasma, and we discussed the possibility that it was an XRB associated with M101 based on its positional coincidence with the end of a spiral arm and short- and long-term variability. However, these properties are also consistent with this source being a background galaxy with a thermal component from e.g. a supernova-driven galactic wind. A further five candidate AGN/background galaxies have also been identified by \citet{wangetal99} based on the colours of optical counterparts, although two of these (\#10 \& 94) are not significantly detected in the hard 2--6\,keV band. In addition, an interarm sources (\#66) with hard X-ray colours has a radio counterpart, which points to it being a background radio galaxy.
  
In total, fifty-four sources are significantly detected ($>4\sigma$) in the hard 2-6\,keV band. From an inspection of the positions of these sources with respect to optical DSS and \hi images, approximately half are coincident with star forming \hii regions or spiral arm structure, arguing that they are associated with M101. The remaining sources are the best candidates for background AGN, and are either interarm or sources located away from the main body of M101. This number is consistent with the number of background sources we expect based on the statistical arguments given above. 

The majority of the hard sources fall into the XRB and absorbed source categories in the X-ray colour classification scheme. It is interesting to note that in Figure~\ref{fig:im_nh} (top), while the thermal SNR candidate sources (plotted in red) correlate well with the \hi structure, many of the absorbed sources (plotted in blue) fall into interarm regions.  The \hi column in M101 ranges from $\sim1\times10^{20} \cms$ in the interarm regions to $\sim10^{22} \cms$ in the spiral arms \citep{braun95}, which implies that interarm absorbed sources with $N_H\ga10^{22} \cms$ must possess absorption in excess of that in the disc of M101, indicating that they are likely to be background galaxies.

We note that the only source in the background source class to be significantly detected in the hard band is \#29. However, apart from the some SSS candidates that lie in this region (see section~\ref{sec:sss}), the remaining source classifications are ambiguous, as they are all borderline between the background source/XRB/SNR categories, with error bars that overlap these categories, and therefore cannot be firmly classified with this data.

\subsection{X-ray binaries (XRBs)}
\label{sec:xrb}

The X-ray colours show that a large proportion of the \xmm sources ($\sim$ 60 per cent) have spectral shapes consistent with XRBs. However, not all of these sources are associated with M101 itself. Again, from an inspection of the source locations with respect to the optical and \hi images, we estimate that approximately half are associated with star-forming \hii regions or spiral arm structure. As shown above in section~\ref{sec:agn}, the majority of background AGN candidates also have X-ray colours in the same spectral range, and constitute the main contaminant in this category. Indeed, both of the optically identified background galaxies (\#1 \& 104), plus two of the five AGN candidates of \citet{wangetal99} fall into this category. In addition, absorbed or nonthermal emission from SNRs may also fall in this spectral range on the X-ray colour diagram. 

Our detailed spectral and timing analyses of the fourteen brightest sources in this field in Paper I allows us to investigate the robustness of this classification scheme. Eleven of the fourteen sources are classed as XRBs in this scheme, which agrees with our interpretation (based on spectral shape and variability) of ten being high-state XRBs, and one being a (now identified) background AGN (\#104/XMM-5). The three exceptions are XMM-7, XMM-11 \& XMM-13, which fall into the thermal SNR range. However, this can be understood because their spectra were generally soft; XMM-7 is a SSS and XMM-11 and XMM-13 both have strong soft components in their spectra. This demonstrates that there is an overlap region between true SNRs and soft XRBs.

Additional evidence that the sources in this category are accreting systems comes from our detection of short-term variability during the \xmm observation. Of the sixteen sources found to have statistical short-term variability at greater than the 95 per cent level, twelve fall into the XRB category. The remaining four fall into the SNR category; however these are all sources separately identified as either stars or soft XRBs where variability on this timescale is expected (see section~\ref{sec:snr}).

\subsection{Supernova remnants (SNRs)}
\label{sec:snr}

SNRs generally have soft thermal X-ray spectra with the majority of their flux below 2\,keV (e.g. \citealt{pietsch04}). Fifteen sources fall into the SNR category in the X-ray colour classification scheme, although four show short-term X-ray variability that we would not expect to detect from SNRs. However, these four are all separately identified either as stars (\#25 \& 105) or bright soft XRBs (\#74/XMM-11 [coincident with SNR MF83] \& \#77/XMM-13). 

Of the eleven non-variable sources in this category, one is a star (\#21), another is a bright soft XRB (\#108/XMM-7) and three are optically identified AGN candidates (\#10, \#85 \& 94, see section~\ref{sec:agn}). Three are unidentified sources (\#32, \#75 \& 84), although two of these (\#32 \& 75) are in in the vicinity of outer spiral arms in areas not covered by the optical survey of \citet{matonick97}, and therefore stand as promising new candidate X-ray SNRs. Further optical/radio observations will be required to confirm their identifications. 

The remaining three sources are consistent with the positions of known SNRs/\hii radio sources. The \xmm position of source \#44 is consistent with that of SN1970G in the giant \hii region NGC~5455, but in the absence of a more accurate \chandra position to confirm this association, it is possible that the X-ray emission we detect with \xmm may be confused with other sources in the \hii region where it resides. However, the \chandra position of source \#67 shows that it is firmly associated with SNR MF65, with positions matching to within $\sim$ 1 arcsecond. In addition, the \chandra position of source \#87 matches one of six radio sources detected by \citet{eck02} in the giant \hii region NGC~5462. The authors do not classify this source as a SNR as it was only detected at 20\,cm and hence has no measured spectral index. Even though we may just be detecting thermal X-ray emission from the \hii region itself, this source does show a factor of $\sim7$ long-term decrease in X-ray luminosity between the \rosat observations in 1992/1996, the \chandra observation in 2000 and the \xmm observation in 2002, giving additional evidence that it may be a cooling SNR. Interestingly, this is one of the two sources that has a significant X-ray colour change between the \chandra and \xmm observations, moving from the XRB to the SNR category with decreasing luminosity. This indicates that this source may be a recent supernova, with the colour change attributable to thermal source cooling. 

In addition, there are two \xmm sources that do not have SNR X-ray colours, but which do have SNR optical counterparts. Source \#40 matches MF22 ($\sim$ 1.7 arcsecond offset), but falls into the XRB category. Although SNRs dominated by nonthermal emission (i.e. Crab-like SNR) will have harder X-ray spectra than thermal SNRs, we must bear in mind that XRBs and SNRs may in fact be spatially coincident, as both are the end products of massive stars and will tend to be located in star-forming regions. The other source (\#46/XMM-1) demonstrates this point, as it is coincident with a SNR (MF37) although the spectral and timing properties of this source presented in Paper I show that it is an XRB.

\subsection{Supersoft sources (SSSs)}
\label{sec:sss}

SSSs have been detected in a number of galaxies (e.g. M31, \citealt{distefano04a}; M81, \citealt{swartz02}; M83, \citealt{soria03}) as well as M101 itself \citep{pence01}. They were classically defined as X-ray sources with effective temperatures of $<$ 100\,eV, consistent with nuclear burning on the surface of white dwarfs accreting at high rates (\citealt{kahabka97}; \citealt{greiner00}). More recently, \citet{distefanokong04} have extended this definition to include sources with either blackbody type spectra with $kT<175$\,eV, or power-law spectra with $\Gamma\geq3.5$, with less than 10 per cent of the total X-ray luminosity above 1.5\,keV. This extended class is likely to represent a heterogeneous group including additional objects such as SNRs, accreting neutron stars or possibly even accreting intermediate-mass black holes \citep{distefano04b}.  For the majority of the sources in this \xmm observation, we do not know the detailed spectral shapes of the sources, but we can identify candidate SSSs if emission is only detected in the softest (0.3--1\,keV) energy band. The X-ray colour diagrams (Figure~\ref{fig:xcolours}) show that 15 sources (marked with stars) fall into this category, with significant detections above 3$\sigma$ {\it only} in the soft band in either the PN and/or MOS data (\#24, 31, 38, 43, 53, 56, 58, 62, 69, 71, 79, 80, 91, 95 \& 100). Simulations in {\small XSPEC} show that a 175\,eV blackbody EPIC (PN+MOS) spectrum would have approximate hardness ratio values of HR1=-0.66 and HR2=-0.98, and the candidate SSSs hardness ratios are consistent with this, although as expected their hard HR2 colours are mostly unconstrained. We can discount source \#56 as a SSS candidate as it is positionally coincident with a foreground star (see section~\ref{sec:stars}), but four sources (\#53, 58, 71 \& 100) are firmly associated with optically identified \hii regions and the remainder generally lie on or near spiral arm structures as shown in the \hi map in Figure~\ref{fig:im_nh} (top). No short-term variability is detected in any of these sources, but long-term variability is evident in sources \#24, 43, 53, 58, 71, 91 \& 100.

\subsection{Foreground stars}
\label{sec:stars}

Four \xmm sources are positionally coincident with GSC stars (\#3, 25, 56 \& 105), and optical counterparts for an additional two (\#21 \& 78) have been identified as stars by \citet{wangetal99}. Stars have soft thermal X-ray spectra, and the X-ray colours of sources \#3, 21 \& 56 are consistent with this, lying at the soft end of the absorbed, SNR and SSS categories respectively. In contrast, source \#78 has a harder X-ray colour equivalent to $\Gamma\sim1.5$ and falls into the XRB category. Given that this source was only detected by the PSPC in the \rosat observations and hence had large positional errors, it is feasible that it was mistakenly associated with a star-like optical counterpart by \citet{wangetal99}. The \xmm position of this source places it in an outer spiral arm of M101 near an \hii region, and the X-ray properties make it more likely that this source is an XRB in M101 itself. 

The remaining two sources are among the brightest X-ray sources in this field (\#25=GSC1275 \& \#105=GSC1069/[AG79] 10), and both have sufficient counts for spectral fitting. We have confirmed their identifications as stars (as opposed to background AGN) by fitting their \xmm spectra in {\small XSPEC v11.3} with a power-law model typically fit to AGN spectra, and {\small MEKAL} thermal plasma models which describe coronal emission from solar-type stars. The brightest source (\#25) has a soft X-ray spectrum, with an X-ray flux of 1.1$\times10^{-13} \ergcms$. It is poorly fit with a power-law model ($\chi^2/dof$=225/114), and the slope of $\Gamma$=6.6 is too steep for most AGN (typically $\Gamma\sim1.7-2.1$, \citealt{nandra94}). It is, however, well fit with a two-temperature plasma ($kT$=0.37/1.01\,keV, $\chi^2/dof$=114/111) with abundances of $\sim$0.2Z$_{\odot}$, consistent with the temperatures and metallicities found in solar-type K-type stars \citep{briggs03}. The light curve of this source also displays a classical flare with an exponential decay (see Figure~\ref{fig:lcurve}, top), often seen in active K stars \citep{briggs03}. The second source (\#105) has an X-ray flux of 4.3$\times10^{-14} \ergcms$, and is similarly well fit with a two-temperature plasma ($kT$=0.27/1.05\,keV, $\chi^2/dof$=36/38) assuming abundances of $\sim$0.3Z$_{\odot}$ (leaving this parameter free resulted in unrealistically low values of $\sim0.1$Z$_{\odot}$). This time, the power-law fit is statistically acceptable ($\chi^2/dof$=46/39), but the slope of $\Gamma$=4.9 is again too steep for the vast majority of AGN. This source is also variable at $>3\sigma$ (see Figure~\ref{fig:lcurve}, bottom).

\section{Summary}
\label{sec:conc}

In this paper we have studied the properties of the X-ray point source population of M101 as seen with {\it XMM-Newton}. We detect 108 X-ray sources within the \d25 ellipse of M101, with fluxes ranging between 3.1$\times10^{-15}$--4.2$\times10^{-13} \ergcms$ corresponding to luminosities of 1.9$\times10^{37}$--2.6$\times10^{39} \ergsec$ in the 0.3--6\,keV energy range at the distance of M101. We have studied the X-ray colour and variability characteristics of these sources, and have supplemented the \xmm dataset with two archival \chandra observations in order to search for flux and spectral changes. The main results can be summarised as follows:

\begin{enumerate}

\item Multiwavelength cross-correlations show that 20 sources are coincident with \hii regions and/or SNRs, 7 have identified/candidate background AGN/galaxy counterparts, 6 are coincident with foreground stars and one has a radio counterpart. The remaining 74 are X-ray detections only.

\item The sources are distributed among the X-ray colour source categories as follows: 62 XRBs, 15 SNRs, 11 absorbed sources, 5 background sources and 15 candidate SSSs (those only detected significantly in the soft 0.3--1\,keV band).

\item We estimate that $\sim$ 24 sources are background sources. Two sources are optically identified as background galaxies, one of which is the bright source XMM-5 from Paper I. A further five candidates background objects have been identified in the \rosat study of M101 \citep{wangetal99}.

\item Approximately 60 per cent of the sources have X-ray colours consistent with XRBs. We estimate that half are associated with \hii regions or spiral arm structure, with background AGN being the main contaminant in this category as they have similar spectral shapes.

\item Fifteen sources have X-ray colours consistent with SNR. Two correlate with the positions of known SNRs, and another with a radio source in the giant \hii region NGC~5462. Two sources are promising new candidate X-ray SNRs, one is unidentified, while the other nine are identified as stars, soft XRBs and AGN candidates.

\item Of the 15 sources with significant detections in the softest X-ray band (0.3--1\,keV), 14 are candidate SSSs; four are firmly associated with \hii regions and the remainder generally lie on or near spiral arm structure. The other source is coincident with a foreground star.

\item Four \xmm sources are coincident with GSC stars, two of which are among the brightest X-ray sources in the field. Optical counterparts for an additional two sources have been identified as stars by \citealt{wangetal99}, although one shows harder X-ray colours than expected for a star and may be an XRB located in a spiral arm. 

\item Sixteen sources are found to be significantly ($>$ 95 per cent) variable during the \xmmn observation. Twelve of these fall into the XRB category, giving additional evidence that they are accreting systems. The remaining four fall into the SNR category; however these are all sources separately identified as either stars or soft XRBs.

\item Using archival \chandra and \rosat data, we find that 44 ($\sim$ 40 per cent) \xmm sources show long-term variability over a baseline of up to $\sim$ 10 years. One source is detected going through a transient phase in the \xmm observation, while comparisons with archival data reveal seven sources exhibiting possible transient behaviour.

\item Of the thirteen sources that vary significantly between the \chandra and \xmm observations (excluding SSSs and foreground stars), only four show the canonical HS to LH state transition seen in Galactic black hole binary candidates, three of which are bright sources studied in Paper I. The remaining sources show the opposite behaviour, though this may simply be a consequence of the X-ray bands used to define the spectral changes.  

\end{enumerate}

\vspace{3mm}

\noindent We have shown that the X-ray colour scheme of \citet{prestwich03}, adapted for the \xmm EPIC instruments, can be a useful guide for source classification, and so can give valuable insights into the overall source population in a spiral galaxy. However, we also demonstrate that there are undoubtedly overlap regions in the colour-colour diagram due to the similar spectral shapes of sources of different natures. For example, soft XRBs may have colours equivalent to thermal SNRs, and background AGN can have colours in the same range as XRB systems. 

This \xmmn observation has provided new information on the X-ray source population of M101, a spiral galaxy similar to the Milky Way.  Using the spectral and timing properties of the brightest sources and the source classifications of the entire \xmm catalogue, we have shown that this X-ray source population ($L_X\sim10^{37}-10^{39} \ergsec$) is dominated by accreting sources (XRBs, SSSs). Further studies, such as the ongoing \chandra/{\it HST} 1\,Ms deep observation program, will extend our knowledge to sources at lower luminosities and allow the identification of multiwavelength counterparts to confirm their identities. The useful methods developed in this study can now be extended to larger samples of local galaxies of different types, to detail and establish the properties of their X-ray source populations, search for statistical trends and characterize their overall X-ray emission properties. This in turn will provide us with the essential tools needed to interpret the X-ray emission detected from galaxies at higher redshifts.

\section*{Acknowledgments}

This work is based on observations obtained with {\it XMM-Newton}, an ESA science mission with instruments and contributions directly funded be ESA and NASA. We thank the referee, Dr K. Mukai, for helpful comments on the original draft of this paper, and we thank K. Briggs for assistance with the source detection techniques. This research has made use of the SIMBAD database, operated at CDS, Strasbourg, France, and the NASA/IPAC Extragalactic Database (NED) which is operated by the Jet Propulsion Laboratory, California Institute of Technology, under contract with the National Aeronautics and Space Administration. The second digitized sky survey was produced by the Space Telescope Science Institute, under Contract No. NAS 5-26555 with the National Aeronautics and Space Administration.  The Guide Star Catalog was produced at the Space Telescope Science Institute under U.S. Government grant. These data are based on photographic data obtained using the Oschin Schmidt Telescope on Palomar Mountain and the UK Schmidt Telescope. LPJ is supported by a PPARC studentship.

\label{lastpage}

{}

\end{document}